 \documentclass[preprint,
                preprintnumbers,
                amsmath,
                amssymb,
                superscriptaddress]{revtex4}

\usepackage{latexsym}
\usepackage{amsfonts}
\usepackage{amsmath}
\usepackage[dvips]{graphicx}
\usepackage{color}
\usepackage{bm}

\newcommand{\placefig}[1]{
    \begin{center}
     {\bf FIGURE \ref{#1} AROUND HERE}
    \end{center}
}

\newcommand{\placetab}[1]{
    \begin{center}
     {\bf TABLE \ref{#1} AROUND HERE}
    \end{center}
}

\def\figfoot{Vendrell et. al., Journal of Chemical Physics}

\newcommand{\figcaption}[2]{
    \noindent {\bf Figure \ref{#1}:} #2
    \vspace{1cm}
}

\def\zun{H$_5$O$_2^+$}
\def\icm{cm$^{-1}$}

\begin{document}

\author{Oriol Vendrell}
\email[e-mail: ]{oriol.vendrell@pci.uni-heidelberg.de}
\affiliation{Theoretische Chemie, Physikalisch-Chemisches Institut, 
   Universit\"at Heidelberg, INF 229,
   D-69120 Heidelberg, Germany}

\author{Fabien Gatti}
\email[e-mail: ]{gatti@univ-montp2.fr}
\affiliation{%
  CTMM, Institut Gerhardt 
(UMR 5232-CNRS), CC 1501, Universit{\'e} de Montpellier II,
  F-34095 Montpellier, Cedex 05, France
}

\author{David Lauvergnat}
\email[e-mail: ]{david.lauvergnat@lcp.u-psud.fr}
\affiliation{CNRS, Laboratoire de Chimie Physique (UMR 8000), Universit\'e
  Paris-Sud, F-91405 Orsay}

\author{Hans-Dieter Meyer}
\email[e-mail: ]{Hans-Dieter.Meyer@pci.uni-heidelberg.de}
\affiliation{Theoretische Chemie, Physikalisch-Chemisches Institut, 
   Universit\"at Heidelberg, INF 229,
   D-69120 Heidelberg, Germany}

\title{Full Dimensional (15D) Quantum-Dynamical Simulation of the Protonated Water-Dimer I:
       Hamiltonian Setup and Analysis of the Ground Vibrational State}

\date{\today}

\begin{abstract}
Quantum-dynamical full-dimensional (15D) calculations are reported for the protonated water dimer ({\zun})
using the multiconfiguration time-dependent Hartree (MCTDH) method.
The dynamics is described by curvilinear coordinates. The expression of the kinetic energy operator
in this set of coordinates is given and its derivation, following the polyspherical method, is discussed.
The PES employed is that of Huang {\em et al.} \mbox{[JCP, \textbf{122}, 044308, (2005)]}.
A scheme for the representation of the potential energy surface (PES) is discussed which is based on a
high dimensional model representation scheme (cut-HDMR), but modified to take advantage of the 
mode-combination representation of the vibrational wavefunction used in MCTDH.
The convergence of the PES expansion used is quantified and evidence is provided that it
correctly reproduces the reference PES at least for the range of energies of interest.
The reported zero point energy of the system is converged with respect to the MCTDH expansion 
and in excellent agreement ($16.7$ {\icm} below) with the diffusion Monte Carlo result on the PES of Huang {\em et al}. 
The highly fluxional nature of the cation is accounted for through use of curvilinear coordinates.
The system is found to interconvert between equivalent minima through wagging and internal rotation motions
already when in the ground vibrational-state, i.e., T=0.
It is shown that a converged quantum-dynamical description of such a flexible, multi-minima system is possible.
\end{abstract}


\maketitle

\section{Introduction}


  The protonated water-dimer {\zun} (Zundel cation) is the smallest system in which an excess proton is shared
  between two water molecules. Much effort has been devoted
  to this system due to the importance that the solvated
  proton has in several areas of chemistry and biology.
  In recent years a fast development of the spectroscopical techniques available to probe the vibrational dynamics
  of ionic species in the gas phase has taken place, and several studies have appeared around the {\zun} 
  system \cite{asm03:1375,fri04:9008,hea04:11523,ham05:244301} and other more 
  complex clusters and molecules \cite{hea05:1765,ros07:249}.
  In order to achieve a satisfactory understanding of the spectroscopy and dynamics
  accurate theoretical results are needed in parallel and several works
  have appeared to fill this gap which are based on different theoretical
  approaches, from classical- to quantum-dynamical 
  methods \cite{ven01:240,dai03:6571,hua05:044308,ham05:244301,sau05:1706,kal06:2933}.
  Accurate quantum-dynamical simulations of the dynamics and
  spectroscopy of the system require of an accurate reference potential energy surface (PES). Very accurate
  PES and dipole moment surfaces (DMS) have been recently produced by 
  Huang et al. \cite{hua05:044308} for {\zun}. They are based on several ten-thousands of coupled-cluster
  calculations combined with a clever fitting algorithm which is based on a redundant set of
  coordinates, namely all atom-atom distances. The PES is able to describe the floppy motions occurring at
  typical energies of excitation in the linear IR regime and dissociates correctly \cite{hua05:044308}.
  Several works have already appeared in which the PES and DMS of Huang et al. \cite{hua05:044308} 
  were used \cite{mc05:061101,ham05:244301,kal06:2933}. In Ref. \cite{mc05:061101}
  full-dimensional vibrational calculations for {\zun} were undertaken using
  diffusion Monte Carlo (DMC) and variational wavefunction methods. The
  zero point energy (ZPE), some vibrational excited-state energies and properties of the ground vibrational
  state of the system are reported and discussed. 
  The PES of Huang et al. was also used in Ref. \cite{ham05:244301} where the experimental
  IR spectrum of the system was reported together with the calculation of relevant excited-vibrational states.

 As will be later discussed, {\zun} features several symmetry-equivalent
 minimum energy structures and large amplitude motions between different regions
 connected by low energy barriers. For such systems curvilinear coordinates, e.g.
 bond-bond angle, dihedral angle, or internal rotation coordinates,  offer an advantage
 over rectilinear coordinates since they provide a physically meaningful description
 of the different large-amplitude molecular motions. As a remark, an attempt was made to describe the
 system by a set of rectilinear coordinates, since they result in a simple expression 
 of the kinetic energy operator (KEO). The resulting Hamiltonian was unsuccessful at describing
 of variuos large-amplitude displacements, which resulted in abandonement of such an approach
 and introduction of a curvilinear-coordinates based Hamiltonian.
 When deriving the KEO for a Hamitonian in curvilinear
 coordinates one may define those in terms of rectilinear atom motions and use
 standard differential calculus to obtain the desired expression
 \cite{pod28:812}. This approach
 becomes extremely complex and error prone already for systems of a few atoms.
 Another possibility is to use an approximate KEO where some cross-terms are
 neglected, thus simplifying the derivation. In this case, however, an
 uncontrolled source of error is introduced in the calculation.
 These drawbacks are overcome by employing the
 polyspherical approach \cite{gat98:8804,gat99:7225,gat01:8275} when defining the coordinate set and
 deriving the corresponding KEO. In the polyspherical method the system is described by a
 set of vectors of any kind (e.g. Jacobi, valence, \ldots). The kinetic energy is given in
 terms of the variation in length and orientation of these vectors,
 the latter is defined by angles with respect to a body fixed frame.  
 Following the polyspherical method the KEO of
 the system at hand is derived in a systematic way without use of
 differential calculus.
 The polyspherical approach has already been discussed in a general framework as well as for
 some molecular systems (see for instance \cite{gat98:8804,gat99:7225,gat01:8275,gat03:507}).  
 In this paper the full derivation of the KEO in
 a set of polyspherical coordinates is illustrated for the {\zun} cation and all
 the key steps are carefully discussed.  The derivation is followed step-by-step
 for this rather complex system, thus providing the basic guidelines to be
 followed for the derivation of KEOs for complex molecular systems and clusters.

  In quantum-dynamical wavefunction-based studies, as the one being introduced here, it
  is convenient to represent the operators on a discrete variable representation (DVR)
  grid, which in the multidimensional case is the direct product of one-dimensional DVRs.
  The potential operator is then given by the value of the PES at each point of the grid.
  In the present case, however, the resulting primitive grid has more than $10^{15}$ points.
  This number makes clear that the potential must be represented in a more compact form
  to make the calculations feasible. Algorithms exist which take advantage of the fact that
  correlation usually involves the concerted evolution of only a small number of
  coordinates as compared to the total number. 
  Hierarchical representations of the PES are then constructed including
  potential-function terms up to a given number of coordinates \cite{bow03:533,rab99:197,li01:1}.
  We discuss here a variation of a hierarchical representation of the PES in which the
  coordinates are grouped together into modes, and the modes are the basic units that
  define the hierarchical expansion. It will be shown that this approach allows for a fast
  convergence of the PES representation if the modes are defined as groups of the most strongly
  correlated coordinates.

  The present work provides new reference results on the properties of the ground vibrational
  state of {\zun}. The convergence of the given results is established by
  comparison to the DMC results \cite{mc05:061101} on the reference PES of Huang
  et al. \cite{hua05:044308}. These results, together with the Hamiltonian defined here,
  set also the theoretical and methodological framework used in the companion paper \cite{ven07:paperII}
  where the IR spectrum and vibrational dynamics of {\zun} are analyzed.



  The paper is organized as follows. Section \ref{sec:MCTDH} presents a brief description
  of the multiconfiguration time-dependent Hartree (MCTDH) method. Section \ref{sec:Hamil}
  details the construction of the Hamiltonian operator for the {\zun} system. The derivation
  of the KEO is discussed in \ref{H5O2+}, while the construction of the potential
  is detailed in \ref{sec:cluster}. Section \ref{sec:ZPE} discusses the calculation of the
  ground vibrational state of the system and gives a comparison to available results. In section
  \ref{sec:wavefunction} some properties of the system in relation to its ground vibrational-state 
  wavefunction are discussed. In section \ref{sec:quality} the quality of the potential
  expansion is analyzed and discussed. Section \ref{sec:conclusions} provides some general
  conclusions.


\section{Brief description of the MCTDH method}  \label{sec:MCTDH}

The \emph{Multiconfiguration time-dependent Hartree} (MCTDH)
method \cite{mey90:73,man92:3199,bec00:1,mey03:251} is a general
algorithm to solve the time-dependent  Schr\"odinger equation.
The MCTDH wave function is expanded in a sum of products
of so--called \emph{single--particle functions} (SPFs).
The SPFs, $\varphi(Q,t)$, may be one-- or multi--dimensional
functions and, in the latter case, the coordinate $Q$ is a collective
one, $Q=(q_k,\cdots,q_l)$. As the SPFs are time--dependent, they
follow the wave packet and often a rather small number of SPFs
suffices for convergence.


The \emph{ansatz} for the MCTDH wavefunction reads
\begin{eqnarray} \label{eq:ansatz}
\Psi(q_1,\cdots,q_f,t) & \equiv & \Psi(Q_1,\cdots,Q_p,t) \nonumber \\
& = & \sum_{j_1}^{n_1} \cdots \sum_{j_p}^{n_p} A_{j_1,\cdots,j_p}(t) \,
\prod_{\kappa=1}^{p} \varphi^{(\kappa)}_{j_\kappa}(Q_\kappa,t) \\
& = & \sum_J A_J \, \Phi_J  \nonumber \; ,
\end{eqnarray}
where $f$ denotes the number of degrees of freedom and $p$ the number
of MCTDH  \emph{particles}, also called \emph{combined modes}.
There are $n_\kappa$ SPFs for the $\kappa$'th particle.
The $A_J \equiv A_{j_1 \ldots j_f}$ denote the MCTDH expansion
coefficients and the configurations, or Hartree-products, $\Phi_J$ are
products of SPFs, implicitly defined by Eq.\ (\ref{eq:ansatz}).
The SPFs are finally represented by linear combinations of
time-independent primitive basis functions or DVR grids.


The MCTDH equations of motion are derived by applying the Dirac-Frenkel
variational principle to the \emph{ansatz}
Eq.\ (\ref{eq:ansatz}). After some algebra, one obtains
\begin{eqnarray}
\label{eq:odea}
i \dot{A}_J & = & \sum_L \, \langle \Phi_J \! \mid \! H \! \mid \!
\Phi_L \rangle \, A_L \; , \\
\label{eq:odephi}
 i \dot{\bm{\varphi}}^{(\kappa)} & = & \left( 1-P^{(\kappa)} \right)
 \left( \bm{\rho}^{(\kappa)}
 \right)^{-1} \! \langle \mathbf{H} \rangle^{(\kappa)}
 \bm{\varphi}^{(\kappa)} \; ,
\end{eqnarray}
where a vector notation, $\bm{\varphi}^{(\kappa)}=(\varphi^{(\kappa)}_1,
\cdots,\varphi^{(\kappa)}_{n_\kappa})^T$, is used.
Details on the derivation,
the definitions of the mean-field $\langle \mathbf{H} \rangle^{(\kappa)}$,
the density $\bm{\rho}^{(\kappa)}$ and the projector $P^{(\kappa)}$, 
as well as more general results, can be
found in Refs.\ \cite{man92:3199,bec00:1,mey03:251}. Here and in the
following (except for Section \ref{sec:Hamil}) we use a unit system with $\hbar=1$.

The MCTDH equations conserve the norm and, for time-independent
Hamiltonians, the total energy. MCTDH simplifies to Time-Dependent
Hartree when setting all $n_\kappa=1$. Increasing the $n_\kappa$
recovers more and more correlation, until finally, when $n_\kappa$
equals the number of primitive functions, the standard method
(i.~e. propagating the wave packet on the primitive basis) is used.
It is important to note, that MCTDH uses variationally optimal SPFs,
because this ensures early convergence.


The solution of the equations of motion requires to build the
mean--fields at every time--step. A fast evaluation of the
mean--fields is hence essential. To this end the
Hamiltonian is written as a sum of products of mono--particle
operators:
\begin{equation}
\label{eq:prod-form}
H = \sum_{r = 1}^s c_r \prod_{\kappa = 1}^p h_r^{(\kappa)} \, ,
\end{equation}
where $h_r^{(\kappa)}$ operates on the $\kappa$-th particle only and
where the $ c_r$ are numbers.
In this case the matrix-elements of the Hamiltonian can be expressed
by a sum of products of mono--mode integrals, the evaluation of
which is fast.
\begin{equation}
\label{eq:prod-form-mat}
\langle\Phi_J|H|\Phi_L\rangle = \sum_{r = 1}^s c_r \prod_{\kappa = 1}^p
\langle\varphi_{j_\kappa}|h_r^{(\kappa)}|\varphi_{l_\kappa}\rangle \, ,
\end{equation}
Similar expressions apply to the matrix of mean-fields 
$\langle \mathbf{H} \rangle^{(\kappa)}$.
For further information on the MCTDH method, 
see http://www.pci.uni-heidelberg.de/tc/usr/mctdh/.

\section{System Hamiltonian}  \label{sec:Hamil}

\subsection{Kinetic energy operator for {\zun}}  \label{H5O2+}

The derivation of the exact
 kinetic energy  operator
for H$_5$O$_2^+$ is based 
on a  polyspherical approach 
which has been devised in previous articles 
\cite{gat98:8804,gat98:8821,iun99:3377,gat99:7225,gat01:8275,gat03:507}.
This approach can be seen as a 
very efficient way to obtain 
kinetic energy operators for the 
family of polyspherical coordinates. 
The formalism can be applied
whatever the number of atoms
and whatever the set of vectors: Jacobi, 
Radau, valence, satellite, etc. 
The formalism is not restricted to total J=0
and hence the operator may include overall rotation
and Coriolis coupling.\\

\begin{figure}[h!]
    \begin{center}
       \includegraphics[width=8.5cm]{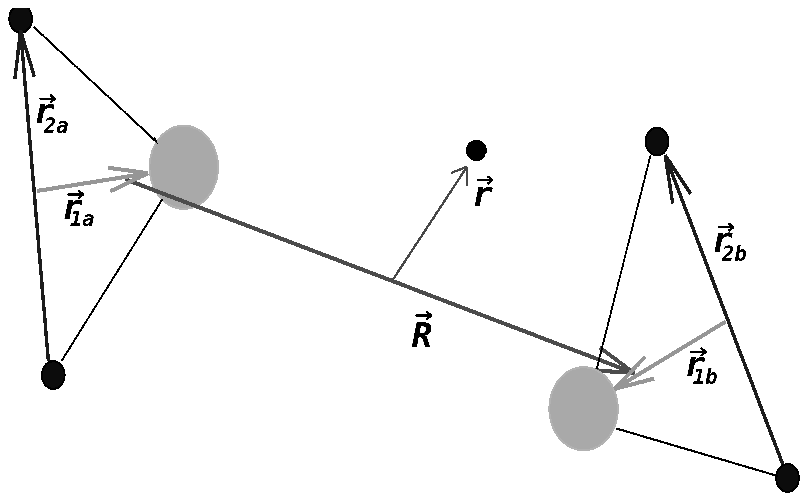}
    \end{center}
    \caption{\figfoot}
    \label{fig:Vect}
\end{figure}

The protonated water dimer system is
described by six Jacobi vectors.
The choice of the Jacobi vectors is not unique
and several clustering schemes are possible, one natural
choice for {\zun} is given in Figure \ref{fig:Vect}.

\placefig{fig:Vect}

The polyspherical approach straightforwardly provides 
the expression of the kinetic energy operator 
in terms of the angular momenta associated
with the vectors describing the system.
Here we use the technique of ``separation into two subsystems" which is
described in Ref.~\onlinecite{gat99:7225} (see Eq. (37) there).
The full kinetic energy operator reads
(we use the notation $\partial_x = \frac{\partial}{\partial x}
$ and $\partial^2_x = \frac{\partial^2}{\partial x^2}$
throughout the paper):

\begin{eqnarray} \label{Eq:48}
&\hat{ T}& = (-\frac{\hbar^2} {2 \mu_R} \frac{1} {R} {\partial^2_R} R) +  
\frac{{({\vec{J}^{\dagger}} \cdot {\vec{J}}+
 {({\vec{L}}_A+{\vec{L}}_B+{\vec{l}})}^2} 
- 2 {  ({\vec{L}}_A+{\vec{L}}_B+{\vec{l}})  
\cdot {\vec{J}})}_{E2}} {2 \mu_R R^2} \nonumber \\
& + & \sum_{i=1}^{2} (-\frac{\hbar^2} {2 \mu_{iA}} \frac{1} {R_{iA}} 
{\partial^2_{R_{iA}}} R_{iA} )  +
  \frac{{({{\vec{L}_A}^2} + 
{\vec{L}_{1A}^{\dagger}} \cdot
 \vec{L}_{1A} - 2 \vec{L}_A \cdot
 {\vec{L}_{1A}})}_{BFA}} {2 \mu_{2A} R_{2A}^2}
 + \frac{{({{\vec{L}_{1A}}^{\dagger}} \cdot
 {\vec{L}_{1A}})}_{BFA}} { 2 \mu_{1A} R_{1A}^2} \nonumber \\
& + & \sum_{i=1}^{2} (-\frac{\hbar^2} {2 \mu_{iB}} \frac{1} {R_{iB}} 
{\partial^2_{R_{iB}}} R_{iB} )  +  \frac{{({{\vec{L}_B}^2} 
+ {{\vec{L}_{1B}}^{ \dagger}} \cdot {\vec{L}_{1B}} - 2 {\vec{L}_B}
\cdot {\vec{L}_{1B}})}_{BFB}} {2 \mu_{2B} R_{2B}^2} 
+ \frac{{({{\vec{L}_{1B}}^{\dagger}} \cdot {\vec{L}_{1B}})}_{BFB}}
 {2 \mu_{1B} R_{1B}^2} \nonumber \\
& - & \frac{\hbar^2}{2m} \frac{1}{r}{\partial^2_{r}} r 
+ \frac{{({\vec{l}}^2)}_{E2}}{2 m r^2} \nonumber \\
\end{eqnarray}

with 

\begin{itemize}
\item R, R$_{1A}$, R$_{2A}$, R$_{1B}$, R$_{2B}$,
r, are the lengths of the vectors 
$\vec{R}$, $\vec{R}_{1A}$, $\vec{R}_{2A}$, $\vec{R}_{1B}$,
 $\vec{R}_{2B}$, $\vec{r}$, respectively.
\item $\mu_{1 \, A (B)}$ = $\frac{m_O 2 m_H}{m_O + 2 m_H}$,
$\mu_{2 \, A (B)}$ = $\frac{m_H}{2}$, $\mu_R$ = 
$\frac{m_O + 2 m_H}{2}$, and $m = \frac{2 m_H  (m_O + 2 m_H)}
{2 m_O + 5 m_H} $. 
\item ${\vec{L}}_{1 \, A (B)} $ is the angular
momentum associated with $\vec{R}_{1 \, A (B)}$.
\item ${\vec{L}}_{2 \, A (B)} $ is the angular
momentum associated with $\vec{R}_{2 \, A (B)}$
and ${\vec{L}}_{A (B)} $ (= ${\vec{L}}_{1 \, A (B)} 
+ {\vec{L}}_{2 \, A (B)}$) is the total angular
momentum associated with monomer A (B).
\item ${\vec{l}}$ is the angular momentum 
associated with the proton.
\item ${\vec{J}}$ is the total angular momentum of the system.
${\vec{J}}$ = ${\vec{l}}$ + ${\vec{L}}_{A} $ 
+ ${\vec{L}}_{B} $ + ${\vec{L}}_{R}$ and
${\vec{L}}_{R}$ is the angular momentum associated 
with $\vec{R}$.
\item E2 is the frame resulting from the two first Euler
rotations starting from the space fixed frame,
 such that the z$^{E2}$ axis lies parallel
to $\vec{R}$. In other words, the two first Euler angles
$\alpha$ and $\beta$
are identical to the two spherical angles of $\vec{R}$
in the space fixed frame.
\item BFA (B) is the body fixed frame associated with 
monomer A (B). 
The two first Euler angles 
$\alpha_{A (B)}$ and $\beta_{A (B)}$ of monomer
A (B) are defined such as z$^{BFA (B)}$ lies parallel 
to $\vec{R}_{2 \, A (B)}$ (in other words they are identical
to the two spherical angles of $\vec{R}_{2 \, A (B)}$
in the E2 frame). The third Euler angle $\gamma_{A (B)}$ is
defined such as $\vec{R}_{1 \, A (B)}$ remains parallel to
the (${(xz)}^{BFA (B)}$, $x^{BFA (B)} > 0$) half plane. 
This definition of BFA(B) is similar to the one recently used to describe
the water dimer (see e.g. Ref \cite{lef02:8710}).
Note however that we have changed the definition of 
the z$^{BFA (B)}$ axis which is now parallel to
the vector between the
two hydrogen atoms and not to the vector joining the center of mass
of H$_2$ to the oxygen atom as in Ref. \cite{lef02:8710}.
This change was performed to avoid the
singularity which appears in the KEO when $\vec{R}$ and
z$^{BFA (B)}$ are parallel.
\end{itemize}

It should be emphasized that all the angular momenta appearing
in the kinetic energy operator are all {\it computed} in the space 
fixed frame but {\it projected} onto the axes of different frames. 
This last point, as well as the properties of 
the corresponding projections, 
is addressed in detail in Ref. \cite{gat03:167}.
We recall here only the main point, i.e. the fact that if a
vector is not involved in the definition of a frame F, the
expression of the projections of the corresponding 
angular momentum onto the F-axes in terms of
the coordinates in this frame, is identical to the usual one in
a space fixed frame. 
This very useful property will be
utilized several times in the following
for instance to obtain Eqs. 
(\ref{Eq:44455ter},\ref{Eq:44455bis},\ref{Eq:44455},\ref{Eq:455},\ref{Eq:proton},\ref{Eq:47}).  \\

We now  slightly change the volume element
 for the lengths of the vectors (as usual),
and assume J=0. The operator can then be recast
 in the following form:

\begin{eqnarray} \label{Eqs:48}
&\hat{ T}& = (-\frac{\hbar^2} {2 \mu_R} {\partial_R^2} ) +
 \frac{  {( {\vec{L}}_A+{\vec{L}}_B+{\vec{l}}  )}^2_{E2} }
 {2 \mu_R R^2} 
-  \frac{\hbar^2}{2m} {\partial_{r}^2}
+ \frac{{{({\vec{l}}^2)}_{E2}}}{2 m r^2}
\nonumber \\
& + & \sum_{i=1}^{2} (-\frac{\hbar^2} {2 \mu_{iA}} 
{\partial_{R_{iA}}^2} )  +  
\frac{{({{\vec{L}_A}^2} +
 {{\vec{L}_{1A}}^{\dagger}} \cdot {\vec{L}_{1A}} - 2 {\vec{L}_A}
\cdot {\vec{L}_{1A}})}_{BFA}} {2 \mu_{2A} R_{2A}^2} +
 \frac{{({{\vec{L}_{1A}}^{\dagger}} 
\cdot {\vec{L}_{1A}})}_{BFA}} { 2 \mu_{1A} R_{1A}^2} \nonumber \\
& + & \sum_{i=1}^{2} (-\frac{\hbar^2} {2 \mu_{iB}}  
{\partial_{R_{iB}}^2} )  + 
 \frac{{({{\vec{L}_B}^2} + {{\vec{L}_{1B}}^{ \dagger}}
\cdot {\vec{L}_{1B}} - 2 {\vec{L}_B}
\cdot {\vec{L}_{1B}})}_{BFB}} {2 \mu_{2B} R_{2B}^2} +
 \frac{{({{\vec{L}_{1B}}^{\dagger}} \cdot {\vec{L}_{1B}})}_{BFB}}
 {2 \mu_{1B} R_{1B}^2} \nonumber \\
\end{eqnarray}

which is to be used with the volume element :

\begin{equation}
dR dR_{1A} dR_{2A} dR_{1B} dR_{2B} dr d\alpha_A \sin \beta_A d\beta_A d\gamma_A
d\alpha_B \sin \beta_B d\beta_B d\gamma_B \sin \theta d\theta d\varphi_H
\sin \theta_{1A} d\theta_{1A} \sin \theta_{1B} d\theta_{1B}
\end{equation}

with

\begin{equation}
\begin{array} {ccccc}
0 &  \leq &  R, R_{1A}, R_{2A}, R_{1B}, R_{2B}, r                  &  < &\infty \nonumber \\
0 &  \leq & \beta_A, \beta_B, \theta, \theta_{1A}, \theta_{1B}     &  \leq &\pi  \nonumber \\
0 &  \leq & \alpha_A, \gamma_A, \alpha_B, \gamma_B, \varphi_H      &  < &2 \pi \nonumber \\
\end{array}
\end{equation}

\begin{figure}[h!]
    \begin{center}
       \includegraphics[width=8.5cm]{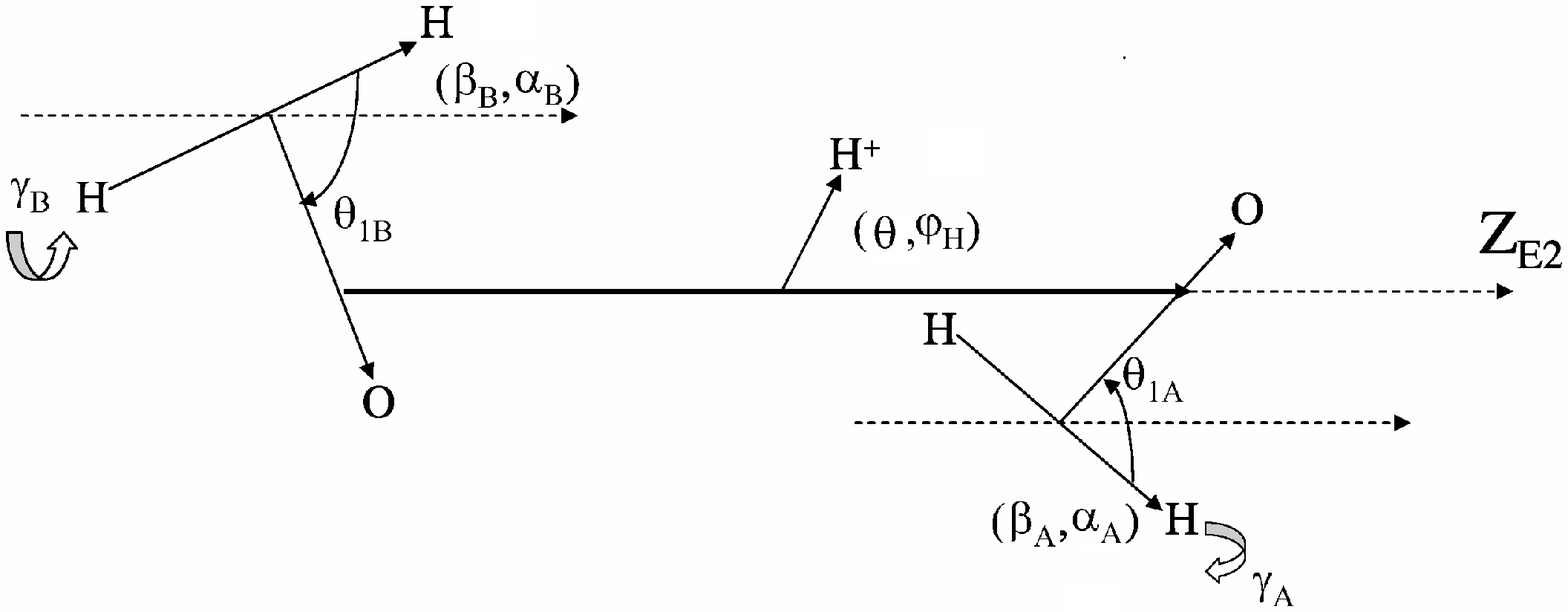}
    \end{center}
    \caption{\figfoot}
    \label{fig:Coord}
\end{figure}

We have 16 degrees of freedom instead of 15 since we have
not defined the third Euler angle yet.
The angles are depicted in Figure \ref{fig:Coord}:

\placefig{fig:Coord}

\begin{itemize}
\item $\theta $ and $\varphi_H$ are the spherical
angles of the proton in the E2 frame: $\theta $
is the angle between z$^{E2}$ and the 
vector $\vec{r}$, $\varphi_H$ describes the rotation
of $\vec{r}$ around z$^{E2}$.
\item $\beta_A$ and $\alpha_A$  are the spherical angles 
of $\vec{R}_{2A}$ in the E2 frame.
The angle $\beta_A$ describes the rocking motion of water A fragment while
$\alpha_A$ describes the rotation of the water A fragment around
the $\vec{R}\equiv\vec{z}_{E2}$ axis (the same for B).
\item $\theta_{1A}$ and $\gamma_A$ are the spherical angles
of $\vec{R}_{1A}$ in the E2A frame, the E2A frame is
the frame obtained after the two Euler rotations  
$\alpha_A$ and $\beta_A$ with respect to the E2 frame.
Consequently, $\theta_{1A}$ is the angle between 
 $\vec{R}_{1A}$  and $\vec{R}_{2A}$ and
$\gamma_A$ describes the rotation of $\vec{R}_{1A}$ 
around $\vec{R}_{2A}$. Hence $\gamma_A$ describes the wagging
motion of the water A fragment (the same for B).
The fact that we have used the separation
into two subsystems for the two water monomers
(see Eq. (37) in Ref. \cite{gat99:7225}) explains why the angles
of $\vec{R}_{1A (B)}$ are defined with respect to
$\vec{R}_{2A (B)}$ and not with respect to $\vec{R}$
as in the standard formulation of the polyspherical
approach (see Eq. (65) in Ref. \cite{gat98:8804}). 
This allows us to have purely intramonomer angles 
instead of angles  mixing the intramonomer and intermonomer 
motions. 
\end{itemize}

Since ${({{\vec{L}_{A (B)}}^2})}_{BFA (B)}$ = ${({{\vec{L}_{A (B)}}^2})}_{E2}$ 
(here we can use ${{\vec{L}_{A (B)}}^2}$ rather
than ${\vec{L}_{A (B)}}^\dagger {\vec{L}_{A (B)}}$ since the projections 
of $\vec{L}_{A (B)}$ onto the axes of the E2 or
BFA (B) frames are hermitian: see Eqs. 
(\ref{Eq:44455ter},\ref{Eq:44455},\ref{Eq:455},\ref{Eq:47}) below),
we can rewrite  the operator as:

\begin{eqnarray} \label{Eq:482}
&\hat{ T}& = (-\frac{\hbar^2} {2 \mu_R}
  {\partial_R^2} )
+ \sum_{i=1}^{2} (-\frac{\hbar^2} {2 \mu_{iA}}  
{\partial_{R_{iA}}^2} ) 
+ \sum_{i=1}^{2} (-\frac{\hbar^2} {2 \mu_{iB}}
  {\partial_{R_{iB}}^2} ) 
-  \frac{\hbar^2}{2m} {\partial_{r}^2}
+ \frac{{{({l}^2)}_{E2}}}{2 m r^2} \nonumber \\
&  + & {({{\vec{L}_A}^2})}_{BFA}
(\frac{1}{2 \mu_R R^2}+\frac{1}{2 \mu_{2A} R_{2A}^2})
+ {({{\vec{L}_B}^2})}_{BFB}
(\frac{1}{2 \mu_R R^2}+\frac{1}{2 \mu_{2B} R_{2B}^2}) \nonumber \\
& + & {({{\vec{L}_{1A}}^{\dagger}} \cdot
{\vec{L}_{1A}})}_{BFA} (\frac{1} {2 \mu_{1A} R_{1A}^2} +
\frac{1} {2 \mu_{2A} R_{2A}^2})
+ {({{\vec{L}_{1B}}^{\dagger}} \cdot {\vec{L}_{1B}})}_{BFB}
 (\frac{1} {2 \mu_{1B} R_{1B}^2} +
\frac{1} {2 \mu_{2B} R_{2B}^2}) \nonumber \\
& - &  \frac{{  ({\vec{L}_A} \cdot
 {\vec{L}_{1A}})}_{BFA}} { \mu_{2A} R_{2A}^2} 
 -   \frac{{  ({\vec{L}_B} \cdot
 {\vec{L}_{1B}})}_{BFB}} { \mu_{2B} R_{2B}^2}
\nonumber \\
&  + & 
\frac{ { ({\vec{L}_A \cdot \vec{L}_B})}_{E2}  }
 { \mu_R R^2}
+ \frac{{(\vec{l}^2)}_{E2}} {2 \mu_R R^2}
+ \frac{  {( \vec{L}_A+\vec{L}_B) \cdot
 \vec{l}_{E2}} }
 { \mu_R R^2}  \nonumber \\ 
\end{eqnarray}

This operator could be straightforwardly used along with
an adequate primitive basis set of spherical harmonics
and Wigner rotation-matrix elements. However, in this work,
we prefer to employ a direct product primitive basis such as
a DVR for all the degrees of freedom (only because the latter
basis set is numerically more efficient for our problem).
We have then to explicitly express of the angular momenta
in terms of the coordinates,
i.e. we have to detail the following terms:

\begin{itemize}
\item (1) ${({{\vec{L}_A}^2})}_{BFA}$, 
${({{\vec{L}_B}^2})}_{BFB}$,
$ {({{\vec{L}_{1A}}^{\dagger}}
\cdot {\vec{L}_{1A}})}_{BFA}$, 
${({{\vec{L}_{1B}}^{\dagger}}
\cdot {\vec{L}_{1B}})}_{BFB}$
\item (2) ${({\vec{L}_A} \cdot
 {\vec{L}_{1A}})}_{BFA}$,  
${({\vec{L}_B} \cdot
 {\vec{L}_{1B}})}_{BFB}$
\item (3) ${({\vec{L}}_A \cdot {\vec{L}}_B)}_{E2}$
\item (4) ${\vec{l}}^2_{E2}$
\item (5) ${({\vec{L}}_A+{\vec{L}}_B)
\cdot {\vec{l}}}_{E2}$
\end{itemize}

but before doing so we have to further specify the coordinates:

\begin{itemize}
\item  (i) to avoid the $1/r^2$ singularity we use the BF Cartesian coordinates $x$, $y$, and $z$ 
for the proton (the definition of BF frame is given in (ii)).
Hence we use, $\partial_{\varphi_H^{BF}} = x \partial_y - y
\partial_x$.
\item (ii) we slightly change the coordinate system (from E2 to BF) and
explicitly define the third Euler angle $\gamma$
of the system and then the BF frame.
We define $\gamma$ such that it is identical to the first
Euler angle of monomer A (here, we slightly break the symmetry
between the two monomers). Consequently, we have:

\begin{eqnarray} \label{Eq:485}
\gamma & =  & \alpha_A  \nonumber \\
\alpha & = & \alpha_B - \alpha_A  \nonumber \\
\varphi_H^{BF} & = & \varphi_H^{E2} - \alpha_A  \nonumber \\
\end{eqnarray}

since J=0, we can then put:

\begin{equation} \label{Eq:486}
\partial_\gamma = 0
\end{equation}

and find

\begin{eqnarray} \label{Eq:487}
\partial_{\alpha_A} & = & -\partial_{\alpha} -\partial_{\varphi_H^{BF}} =
-\partial_{\alpha} - (x \partial_{y} -y \partial_{x}) \nonumber \\
\partial_{\alpha_B} & = & \partial_{\alpha} \nonumber \\
\end{eqnarray}
Note that this transformation does not affect the 
other angles.

\item (iii) in order to have hermitian conjugate momenta
for the angles, we use u$_{\beta_A}$ = $ \cos{\beta_A}$,
u$_{\beta_B}$ = $ \cos{\beta_B}$,
u$_{\theta_{1A}}$ = $\cos{ \theta_{1A}  }$, 
u$_{\theta_{1B}}$ = $\cos{ \theta_{1B}  }$.\\
For each angle $\eta$, we have (with u$_\eta$ =
$\cos \eta$):
 $\partial_{\eta} = - \sin \eta \partial_{u_\eta}
 = - \sqrt{1-u_\eta^2} \partial_{u_\eta}$
and $\partial_{\eta}^\dagger = - \partial_{u_\eta} \sqrt{1-u_\eta^2}$.
\end{itemize}

We have now 15 degrees of freedom:
\begin{equation}                        
 \begin{array} {ccccc}   
  0  & \leq &  R, R_{1A}, R_{2A}, R_{1B}, R_{2B}                         & < & \infty \nonumber \\      
 -1  & \leq & u_{\beta_A}, u_{\beta_B}, u_{\theta_{1A}}, u_{\theta_{1B}} & \leq & 1  \nonumber \\
  0  & \leq & \alpha, \gamma_A, \gamma_B                                 & < & 2 \pi \nonumber \\  
- \infty & < & x, y, z                                                & < & \infty \nonumber \\
  \end{array}
 \end{equation}
Let us now detail all the terms appearing in the kinetic
energy operator:

\begin{itemize}
\item (1) For  $\vec{L}_{B \, BFB}$, the situation is simple
since monomer B is not involved in the definition
of the body fixed frame of the whole system.
Thus, we have :
\begin{equation} \label{Eq:44455ter}
 \begin{array}  {l}
 {{L}}_{B \,   x^{BFB}}  =
 i \hbar \frac{\cos{\gamma_B}} {\sin \beta_B}
  { \partial_{\alpha_B}}
+ i \hbar  \sin \gamma_B \sin \beta_B
  { \partial_{u_{\beta B}}}
 -i \hbar \frac{u_{\beta B}} {\sin \beta_B} { \cos
\gamma_B}
 { \partial_{{\gamma}_B}}\\
{{L}}_{B \, y^{BFB}} =  -i \hbar \frac{\sin \gamma_B}
 {\sin \beta_B}  { \partial_{\alpha_B}}
 + i \hbar \cos \gamma_B \sin \beta_B  { \partial_{u_{\beta B}} }
+ i \hbar \frac{u_{\beta B}} {\sin \beta_B} \sin \gamma_B
  { \partial_{{\gamma}_B}} \\
 {{L}}_{B \,  z^{BFB}} =  -i \hbar
  { \partial_{\gamma_B}}
\end{array}
\end{equation}
and
\begin{eqnarray}\label{Eq:44455bis}
{{\vec{L}_{B\;BFB}}^2} & = &
 - \hbar^2  (\partial_{u_{\beta_B}} {(1-u_{\beta_B}^2)} \partial_{u_{\beta_B}}
+ \frac{1} {1-u_{\beta_B}^2} (\partial_{\alpha_B}^2 +
\partial_{\gamma_B}^2 - 2 u_{\beta_B} \partial_{\alpha_B} \partial_{\gamma_B}))
\end{eqnarray}

Similarly ${{\vec{L}_{A\;BFA}}^2}$
 can be seen
as the total angular momentum of monomer A projected  onto
the axes of the Body fixed frame of the monomer. We obtain the 
usual expression:

\begin{equation} \label{Eq:44455}
 \begin{array}  {l}
 {{L}}_{A \,   x^{BFA}}  =
 i \hbar \frac{\cos{\gamma_A}} {\sin \beta_A}
  { \partial_{\alpha_A}}
+ i \hbar  \sin \gamma_A \sin \beta_A 
  { \partial_{u_{\beta A}}}
 -i \hbar \frac{u_{\beta A}} {\sin \beta_A} { \cos
\gamma_A} 
 { \partial_{{\gamma}_A}}\\
{{L}}_{A \, y^{BFA}} =  -i \hbar \frac{\sin \gamma_A}
 {\sin \beta_A}  { \partial_{\alpha_A}}
 + i \hbar \cos \gamma_A \sin \beta_A  { \partial_{u_{\beta A}} }
+ i \hbar \frac{u_{\beta A}} {\sin \beta_A} \sin \gamma_A
  { \partial_{{\gamma}_A}} \\
 {{L}}_{A \,  z^{BFA}} =  -i \hbar
  { \partial_{\gamma_A}}
\end{array}
\end{equation}
and
\begin{eqnarray}\label{Eqs:1}
{\vec{L}_{A\;BFA}}^2 & = &   -  \hbar^2 
 (\partial_{u_{\beta_A}} {(1-u_{\beta_A}^2)} \partial_{u_{\beta_A}}
+ \frac{1} {1-u_{\beta_A}^2} (\partial_{\alpha_A}^2 +
\partial_{\gamma_A}^2 - 2 u_{\beta_A} \partial_{\alpha_A} \partial_{\gamma_A}))
\end{eqnarray}
However, since monomer A is involved in the definition of the
third Euler angle of the system, we apply the change of 
coordinates, Eqs. (\ref{Eq:485}) and
 (\ref{Eq:487}):
\begin{eqnarray} \label{Eq:4871}
(\partial_{\alpha_A}^2 & + &
\partial_{\gamma_A}^2 - 2 u_{\beta_A} \partial_{\alpha_A} \partial_{\gamma_A})
= \nonumber \\
(\partial_{\alpha}^2 & + & x^2 \partial_y^2 
+ y^2 \partial_x^2 -x \partial_x \partial_y y
 -y \partial_y \partial_x x \nonumber \\
& + & 2 x \partial_y \partial_{\alpha} - 2 y \partial_x \partial_{\alpha}
+ \partial_{\gamma_A}^2 + 2 u_{\beta_A} \partial_{\alpha} \partial_{\gamma_A}
\nonumber \\
& + & 2 u_{\beta_A} \partial_{\gamma_A} x \partial_y -
 2 u_{\beta_A} \partial_{\gamma_A}
y \partial_x) \nonumber \\
\end{eqnarray}

For the projections of $\vec{R}_{1 A (B)}$ onto to
the BFA (B) axes, the situation is not straightforward since
$\vec{R}_{1 A (B)}$ is involved in the definition of BFA (B).
However, since we have used the same convention for the
BF frames as in our previous articles, the expressions of these
projections is already known (identical to Eq. (47a) in
Ref. \cite{gat98:8804} for instance):

\begin{equation} \label{Eq:4455}
 \begin{array}  {l}
 {L}_{1 A \,  x^{BFA}}  
=
i \hbar \frac{u_{\theta 1 A}} {\sin{\theta_{1A}}}
  { \partial_{\gamma_A}}
\\
{L}_{1 A \,  y^{BFA}} =   i \hbar
\sin{\theta_{1A}}   {\partial_{u_{\theta 1 A}}}
\\
 {L}_{1 A \,  z^{BFA}} =  -i \hbar 
{ \partial_{\gamma_A}}
\end{array}
\end{equation}

(This expression can be obtained starting from the E2A
frame which is the frame resulting from the two first Euler 
rotations $\gamma_A$ and $\beta_A$.
The expression of the projections of $\vec{L}_{1 A}$
onto this E2A frame are the usual ones since
$\vec{R}_{1 \, A}$ is not involved in the definition
of this frame. Applying then the third Euler rotation
and the coordinate transformation leading to the
BFA coordinates yields Eq. (\ref{Eq:4455})).
Note that ${L}_{1 A \,  y^{BFA}}$ is not
hermitian (this is due to the fact that $\vec{R}_{1 A}$ is
involved in the definition of BFA).
However, at the end we obtain the usual expression:

\begin{eqnarray}
{({{\vec{L}_{1A}}^{\dagger}}
\cdot {\vec{L}_{1A}})}_{BFA}
& = & - \hbar^2  (\partial_{u_{\theta{1A}}} {(1-u_{\theta_{1A}}^2)}
\partial_{u_{\theta{1A}}} + \frac{1} 
{1-u^2_{\theta_{1A}}} \partial_{\gamma_A}^2)
\end{eqnarray}

and exactly the same for B:

\begin{eqnarray}
{({{\vec{L}_{1B}}^{\dagger}}
\cdot {\vec{L}_{1B}})}_{BFB}
& = & - \hbar^2  (\partial_{u_{\theta{1B}}} {(1-u_{\theta_{1B}}^2)}
\partial_{u_{\theta{1B}}} + 
\frac{1} {1-u^2_{\theta_{1B}}} \partial_{\gamma_B}^2) 
\end{eqnarray}
\item (2) The terms in ${({\vec{L}_{A (B)}}  \cdot
{\vec{L}_{1A (B)}})}_{BFA (B)}$ 
are not obvious since $\vec{R}_{1 \, A (B)}$ is involved
in the definition of BFA (B).
First, let us explicitly highlight the hermiticity
of the term:

\begin{eqnarray} \label{Eq:491}
{({\vec{L}_A} \cdot {\vec{L}_{1A}})}_{BFA} & = &
1/2 {({\vec{L}_A}^\dagger \cdot {\vec{L}_{1A}} +
{\vec{L}_{1A}}^\dagger \cdot {\vec{L}_{A}})_{BFA}} \nonumber \\
\end{eqnarray}
Combining  Eq. (\ref{Eq:4455}) and Eq. (\ref{Eq:44455}) (as
well as Eq. (\ref{Eq:485}) and Eq. (\ref{Eq:487})), we obtain:
\begin{eqnarray} \label{Eq:493}
 & & {({\vec{L}_A} \cdot {\vec{L}_{1A}})}_{BFA}  =  \nonumber \\
& - &  \frac{\hbar^2}{2} [- (\partial_{\alpha} + 
x \partial_y - y \partial_x)
\frac{u_{\theta_{1A}}}{\sin \theta_{1A}}
\frac{\cos \gamma_A}{\sin \beta_A} \partial_{\gamma_A} \nonumber \\
 & + & \partial_{u_{\beta_A}} \frac{u_{\theta_{1A}}}{\sin \theta_{1A}}
 \sin \beta_A  \sin \gamma_A \partial_{\gamma_A}
- 2 \partial_{\gamma_A} \frac{u_{\beta_A}}{\sin \beta_A}
\cos \gamma_A \frac{u_{\theta_{1A}}}{\sin \theta_{1A}} \partial_{\gamma_A}
\nonumber \\
 & + & 2 \partial_{\gamma_A}^2  
+ (\partial_{\alpha} + x \partial_y - y
\partial_x)
\frac{\sin \gamma_A}{\sin \beta_A} \sin \theta_{1A} \partial_{u_{\theta_{1A}}}
\nonumber \\
 & + & \partial_{u_{\beta_A}} \sin \beta_A \cos \gamma_A
\sin \theta_{1A} \partial_{u_{\theta_{1A}}}
+ \partial_{\gamma_A} \frac{u_{\beta_A}}{\sin \beta_A}
\sin \gamma_A \sin \theta_{1A} \partial_{u_{\theta_{1A}}} \nonumber \\
& - & \partial_{\gamma_A} \frac{u_{\theta_{1A}}}{\sin \theta_{1A}}
\frac{\cos \gamma_A}{\sin \beta_A}
(\partial_{\alpha} + x \partial_y - y
\partial_x)
+ \partial_{\gamma_A} \frac{u_{\theta_{1A}}}{\sin \theta_{1A}}
\sin \beta_A \sin \gamma_A \partial_{u_{\beta_A}} \\
& + & \partial_{u_{\theta_{1A}}}
\frac{\sin \gamma_A}{\sin \beta_A} \sin \theta_{1A}
(\partial_{\alpha}  + x \partial_y - y
\partial_x) \nonumber \\
& + & \partial_{u_{\theta_{1A}}}
\sin \beta_A \cos \gamma_A \sin \theta_{1A}
\partial_{u_{\beta_A}} + \partial_{u_{\theta_{1A}}}
\frac{u_{\beta_A}}{\sin \beta_A}
\sin \gamma_A \sin \theta_{1A} \partial_{\gamma_A}] \nonumber \\
\end{eqnarray}
and for B (the situation is a little bit simpler since
monomer B is not involved in the definition of the third 
Euler angle see Eq.  (\ref{Eq:487})):
\begin{eqnarray} \label{Eq:492}
 & & {({\vec{L}_B} \cdot {\vec{L}_{1B}})}_{BFB}  =  \nonumber \\
& - & \frac{\hbar^2}{2} [\partial_{\alpha} \frac{u_{\theta_{1B}}}{\sin \theta_{1B}}
 \frac{\cos \gamma_B}{\sin \beta_B}
\partial_{\gamma_B}
+ \partial_{u_{\beta_B}} \frac{u_{\theta_{1B}}}{\sin \theta_{1B}}
 \sin \beta_B  \sin \gamma_B \partial_{\gamma_B}
- 2 \partial_{\gamma_B} \frac{u_{\theta_{1B}}}{\sin \theta_{1B}}
\cos \gamma_B \frac{u_{\beta_B}}{\sin \beta_B}
\partial_{\gamma_B} \nonumber \\
 & + & 2 \partial_{\gamma_B}^2  - \partial_{\alpha}
\frac{\sin \gamma_B}{\sin \beta_B} \sin \theta_{1B}
\partial_{u_{\theta_{1B}}} + \partial_{u_{\beta_B}} \sin \beta_B \cos \gamma_B
\sin \theta_{1B} \partial_{u_{\theta_{1B}}}
+ \partial_{\gamma_B} \frac{u_{\beta_B}}{\sin \beta_B}
\sin \gamma_B \sin \theta_{1B} \partial_{u_{\theta_{1B}}} \nonumber \\
& + & \partial_{\gamma_B} \frac{u_{\theta_{1B}}}{\sin \theta_{1B}}
\frac{\cos \gamma_B}{\sin \beta_B} \partial_{\alpha}
+ \partial_{\gamma_B} \frac{u_{\theta_{1B}}}{\sin \theta_{1B}}
 \sin \beta_B \sin \gamma_B \partial_{u_{\beta_B}} \nonumber \\
& - & \partial_{u_{\theta_{1B}}}
\frac{\sin \gamma_B}{\sin \beta_B} \sin \theta_{1B}
\partial_{\alpha} + \partial_{u_{\theta_{1B}}}
\sin \beta_B \cos \gamma_B \sin \theta_{1B}
\partial_{u_{\beta_B}}   \nonumber \\
& + & \partial_{u_{\theta_{1B}}}
\frac{u_{\beta_B}}{\sin \beta_B}
\sin \gamma_B \sin \theta_{1B}  \partial_{\gamma_B}] \nonumber \\
\end{eqnarray}
\item (3) For the term in 
$1/2 {(  {\vec{L}}_A^\dagger \cdot {\vec{L}}_B )}_{E2}$,
we note that the expression of the projections 
of ${\vec{L}}_{A (B)}$ onto the E2 can be seen as 
those of total angular momentum of a system onto the axes
of the Space fixed frame. 
Indeed, since monomers A and B are not involved in the
definition of E2, we have  (similar to
Eq. (18) in Ref. \cite{gat98:8804}):

\begin{equation} \label{Eq:455}
 \begin{array}  {l}
 {{L}}_{A \,  x^{E2}}  =
i \hbar \cos \alpha_A \frac{u_A} {\sin \beta_A}  { \partial_{\alpha_A}}
- i \hbar \sin \alpha_A  \sin \beta_A { \partial_{u_{\beta A}} }
 -i \hbar \frac{ \cos
\alpha_A} { \sin \beta_A} { \partial_{{\gamma}_A}}\\
{{L}}_{A \,  y^{E2}} =  i \hbar \sin \alpha_A
 \frac{u_A} {\sin \beta_A}  { \partial_{\alpha_A}}
 + i \hbar \cos \alpha_A \sin \beta_A  {\partial_{u_{\beta A}} }
- i \hbar \frac{ \sin \alpha_A} { \sin \beta_A}  {\partial_{{\gamma}_A}} \\
 {{L}}_{A \, z^{E2}} =  -i \hbar 
{\partial_{\alpha_A}}
\end{array}
\end{equation}

After, noticing that
\begin{eqnarray} \label{Eq:484}
 {( {\vec{L}}_A \cdot {\vec{L}}_B)}_{E2} & =  &
\frac{1}{2} {(  {\vec{L}}_A^\dagger \cdot {\vec{L}}_B +
 {\vec{L}}_B^\dagger \cdot {\vec{L}}_A  )}_{E2} \nonumber \\
\end{eqnarray}

we obtain after using Eq. (\ref{Eq:487}):

\begin{eqnarray} \label{Eq:490}
 & & \frac{1}{2} {(  {\vec{L}}_A^\dagger \cdot {\vec{L}}_B )}_{E2} =
 -  \frac{\hbar^2}{2}  [ - \partial_{\alpha}
\cos{\alpha} \frac{u_{\beta_A}}{\sin \beta_A}
\frac{u_{\beta_B}}{\sin \beta_B} \partial_{\alpha} \nonumber \\
& - & (x \partial_y - y
\partial_x) 
\cos{\alpha} \frac{u_{\beta_A}}{\sin \beta_A}
\frac{u_{\beta_B}}{\sin \beta_B} \partial_{\alpha} \nonumber \\
& + & \partial_{\gamma_A} \frac{\cos{\alpha}}{\sin \beta_A \sin \beta_B}
\partial_{\gamma_B}
 +  \partial_{\alpha} \sin{\alpha} \frac{u_{\beta_A}}{\sin \beta_A}
\sin \beta_B \partial_{u_{\beta_B}} 
+ (x \partial_y - y
\partial_x)   
\sin{\alpha} \frac{u_{\beta_A}}{\sin \beta_A}
\sin \beta_B \partial_{u_{\beta_B}}
\nonumber \\
& + & \partial_{\alpha} \cos{\alpha} \frac{u_{\beta_A}}{\sin \beta_A}
\frac{1}{\sin \beta_B} \partial_{\gamma_B}
+  \partial_{u_{\beta_A}} \sin{\alpha} \sin \beta_A
\frac{u_{\beta_B}}{\sin \beta_B} \partial_{\alpha} \nonumber \\
& + & (x \partial_y - y
\partial_x) \cos{\alpha} \frac{u_{\beta_A}}{\sin \beta_A}
\frac{1}{\sin \beta_B} \partial_{\gamma_B} \nonumber \\
& + & \partial_{u_{\beta_A}} \cos{\alpha} \sin \beta_A
\sin \beta_B \partial_{u_{\beta_B}}
 -  \partial_{u_{\beta_A}} \sin{\alpha}
\frac{\sin \beta_A}{\sin \beta_B} \partial_{\gamma_B} \nonumber \\
& - & \partial_{\gamma_A} \cos{\alpha} \frac{1}{\sin \beta_A}
\frac{u_{\beta_B}}{\sin \beta_B} \partial_{\alpha}
 +  \partial_{\gamma_A} \sin{\alpha}
\frac{\sin \beta_B}{\sin \beta_A} \partial_{u_{\beta_B}}
- \partial^2_{\alpha}
- \partial_{\alpha} (x \partial_y - y
\partial_x) ] \nonumber \\
\end{eqnarray}

\item (4) Since the proton is not involved in the definition
of the whole body fixed frame, the expression
of the projections of $\hat{\vec{l}}$  onto the E$_2$ (or BF)
axes is the usual one. At the end, we obtain:

\begin{eqnarray}\label{Eq:proton}
{({\vec{l}})}^2_{E2} & = & {({\vec{l}})}^2_{BF} =
- \hbar^2
(y^2 \partial_z^2 + z^2 \partial_y^2 
+ z^2 \partial_x^2 + x^2 \partial_z^2
 + x^2 \partial_y^2 + y^2 \partial_x^2 \nonumber \\
 & & -y \partial_y \partial_z z - \partial_y
 y z \partial_z - \partial_x x z \partial_z 
-x \partial_x  \partial_z z - x \partial_x \partial_y y 
- \partial_x x y \partial_y)
\end{eqnarray}

\item (5)
For the last term ${({\vec{L}}_A+{\vec{L}}_B)
\cdot {\vec{l}}}_{E2}
 = 
  {{({\vec{L}}_A+{\vec{L}}_B)^\dagger  \cdot {\vec{l}} +
{\vec{l}}^\dagger \cdot ({\vec{L}}_A+{\vec{L}}_B)}_{BF}}$, 
we need to know the expression of the projections of
${\vec{L}}_{A (B)}$ onto the body fixed axes.
For ${\vec{L}}_{B}$, there is no particular problem
since B is not involved in the definition of $\gamma$
and thus of the BF frame:

\begin{equation} \label{Eq:47}
  \begin{array}  {l}
 {{L}}_{B \,  x^{BF}}  =  i \hbar
\frac{u_{\beta B}}{\sin \beta_B} \cos \alpha  {\partial_{\alpha}}
- i \hbar \sin \alpha \sin \beta_B  { \partial_{u_{\beta B}}}
 - i \hbar \frac{\cos \alpha}{\sin \beta_{\beta B}} { \partial_{{\gamma}_B}}\\
 {{L}}_{B \,  y^{BF}} = i \hbar \sin \alpha \frac{u_{\beta B}} {\sin \beta_B}
 { \partial_{\alpha}}
+i \hbar \cos \alpha \sin \beta_B  { \partial_{u_{\beta B}}}
-i \hbar \frac{\sin \alpha}{\sin \beta_B} {\partial_{{\gamma}_B}} \\
  {{L}}_{B \,  z^{BF}} =  -i \hbar 
{ \partial_{\alpha}}
\end{array}
\end{equation}

For ${\vec{L}}_{A}$, it is less straightforward.
However, applying the third Euler rotation to Eq. (\ref{Eq:455})
and using Eq. (\ref{Eq:487}) yields: 

\begin{equation} \label{Eq:466}
 \begin{array}  {l}
{{L}}_{A \,  x^{BF}}  =
- i \frac{u_{\beta A}}{\sin \beta_A} ( {\partial_{\alpha}}
+  x \partial_{y} -y \partial_{x})
 - \frac{i}{\sin \beta_A}  {\partial_{{\gamma}_A}}\\
 {{L}}_{A \,  y^{BF}} = i \sin \beta_A  
{ \partial_{u_{\beta A}}} \\
  {{L}}_{A \,  z^{BF}} =  i ({\partial_{\alpha}}
+ x \partial_{y} -y \partial_{x})
\end{array}
\end{equation}
and thus:
\begin{eqnarray}
{({\vec{L}}_A+{\vec{L}}_B) \cdot {\vec{l}}}_{E2}
& = & \frac{1}{2} 
  {({({\vec{L}}_A+{\vec{L}}_B)^\dagger \cdot {\vec{l}} +
{\vec{l}}^\dagger \cdot ({\vec{L}}_A+{\vec{L}}_B))}_{BF}} 
 =     \frac{\hbar^2} {2} [  (-
2 \frac{u_{\beta A}}{\sin \beta_A} {\partial_{\alpha}}
 - 2 \frac{1}{\sin \beta_A} {\partial_{{\gamma}_A}}
\nonumber \\
& + & \frac{u_{\beta B}}{\sin \beta_B} \cos \alpha   
{\partial_{\alpha}}
+ { \partial
 \alpha} \frac{u_{\beta B}}{\sin \beta_B} \cos \alpha
- \sin \alpha \sin \beta_B {\partial_{u_{\beta B}}}
\nonumber \\
& - &  {\partial_{u_{\beta B}}} \sin \alpha \sin \beta_B
 - 2 \frac{\cos \alpha}{\sin \beta_B} 
{\partial_{{\gamma}_B}}
) (y { \partial_z} - z  { \partial_y }) \nonumber \\
& + & ( \sin \beta_A  { \partial_{u_A}}
+  {\partial_{u_{\beta A}}}  \sin \beta_A
\nonumber \\
& + & \sin \alpha \frac{u_{\beta B}}{\sin \beta_B}  {\partial_{\alpha}}
+  {\partial_{\alpha}} \sin \alpha \frac{u_B}{\sin \beta_B}
+ \cos \alpha \sin \beta_B  { \partial_{u_{\beta B}}}
\nonumber \\
& + &  { \partial_{u_{\beta B}}} \cos \alpha \sin \beta_B
- 2 \frac{\sin \alpha} {\sin \beta_B}
 { \partial_{{\gamma}_B}}) (z { \partial_x} - x 
{ \partial_z} )\nonumber \\
 & + & 2 (x^2 \partial_y^2 + y^2  \partial_x^2 -
 x \partial_x  \partial_y y -  \partial_x x y \partial_y)] \nonumber \\
&-& {\hbar^2} [ \frac{u_A}{\sin \beta_A}
(x \partial_y y \partial_z + x y \partial_y \partial_z
 -2 x \partial_y^2 z -2  \partial_x y^2  \partial_z 
+ \partial_x y \partial_y z + \partial_x \partial_y y z)]
\nonumber \\
\end{eqnarray}

\end{itemize}

The final expression of the operator in Eq. (\ref{Eq:482}) is recast as 
$\hat{T} = \hat{T}_1 + \hat{T}_2 + \hat{T}_3 + \hat{T}_4$
with 

\begin{eqnarray} \label{Eqs:482}
&\hat{ T}_1& = (-\frac{\hbar^2} {2 \mu_R}
  \frac{\partial^2}{\partial_R^2} )
+ \sum_{i=1}^{2} (-\frac{\hbar^2} {2 \mu_{iA}}
\frac{\partial^2 }{\partial_{R_{iA}}^2} )
+ \sum_{i=1}^{2} (-\frac{\hbar^2} {2 \mu_{iB}}
  \frac{\partial^2 }{\partial_{R_{iB}}^2} )
-  \frac{\hbar^2}{2m} \frac{\partial^2}{\partial_{r}^2}
+ \frac{{{({l}^2)}_{E2}}}{2 m r^2} \nonumber \\
&  + & {({{\vec{L}_A}^2})}_{BFA}
(\frac{1}{2 \mu_R R^2}+\frac{1}{2 \mu_{2A} R_{2A}^2})
+ {({{\vec{L}_B}^2})}_{BFB}
(\frac{1}{2 \mu_R R^2}+\frac{1}{2 \mu_{2B} R_{2B}^2}) \nonumber \\
& + & {({{\vec{L}_{1A}}^{\dagger}} \cdot
{\vec{L}_{1A}})}_{BFA} (\frac{1} {2 \mu_{1A} R_{1A}^2} +
\frac{1} {2 \mu_{2A} R_{2A}^2})
+ {({{\vec{L}_{1B}}^{\dagger}} \cdot {\vec{L}_{1B}})}_{BFB}
 (\frac{1} {2 \mu_{1B} R_{1B}^2} +
\frac{1} {2 \mu_{2B} R_{2B}^2}) \nonumber \\
\end{eqnarray}

\begin{eqnarray} \label{Eqs:4821}
&\hat{ T}_2 &  = 
 -   \frac{{  ({\vec{L}_A} \cdot
 {\vec{L}_{1A}})}_{BFA}} { \mu_{2A} R_{2A}^2}
 -   \frac{{  ({\vec{L}_B} \cdot
 {\vec{L}_{1B}})}_{BFB}} { \mu_{2B} R_{2B}^2}
\nonumber \\
\end{eqnarray}

\begin{eqnarray} \label{Eqs:4822}
&\hat{ T}_3 &  = 
\frac{ { ({{\vec{L}}_A \cdot {\vec{L}}_B})}_{E2}  }
 { \mu_R R^2}
\end{eqnarray}

\begin{eqnarray} \label{Eqs:4823}
&\hat{ T}_4 &  =
 \frac{{({\vec{l}}^2)}_{E2}} {2 \mu_R}
+ \frac{  {( {\vec{L}}_A + {\vec{L}}_B) 
\cdot {\vec{l}}}_{E2} }
 { \mu_R R^2}  \nonumber \\
\end{eqnarray}

The expression of the operator
in terms of the 15 degrees of freedom:
$R, R_{1A}, R_{2A}, R_{1B}, R_{2B}, x, y ,z, \alpha, u_{\beta_A},
 \gamma_A,
u_{\beta_B}, \gamma_B, 
u_{\theta_{1A}}, u_{\theta_{1B}}$
 reads:
(note that the hermiticity clearly appears)
 
\begin{eqnarray} \label{Eq:494}
&\hat{ T}_1 & = (-\frac{\hbar^2} {2 \mu_R}  {\partial_R^2} )
+ \sum_{i=1}^{2} (-\frac{\hbar^2} {2 \mu_{iA}}  {\partial_{R_{iA}}^2} )
+ \sum_{i=1}^{2} (-\frac{\hbar^2} {2 \mu_{iB}}  {\partial_{R_{iB}}^2} )
  -  \frac{\hbar^2}{2m} ({\partial_{x}^2}
+ {\partial_{y}^2} +
{\partial_{z}^2})
\nonumber \\
&  - & \hbar^2  (\partial_{u_{\beta_A}} {(1-u_{\beta_A}^2)} \partial_{u_{\beta_A}}
+ \frac{1} {1-u_{\beta_A}^2} (
\partial_{\alpha}^2  +  x^2 \partial_y^2 + y^2  
\partial_x^2 -x \partial_x \partial_y y
 -y \partial_y \partial_x x \nonumber \\
& + & 2 x \partial_y \partial_{\alpha} - 2 y \partial_x \partial_{\alpha}
+ \partial_{\gamma_A}^2 + 2 u_{\beta_A} \partial_{\alpha} \partial_{\gamma_A}
\nonumber \\
& + & 2 u_{\beta_A} \partial_{\gamma_A} x \partial_y
 - 2 u_{\beta_A} \partial_{\gamma_A}
y \partial_x))
(\frac{1}{2 \mu_R R^2}+\frac{1}{2 \mu_{2A} R_{2A}^2}) \nonumber \\
&  - & \hbar^2  (\partial_{u_{\beta_B}} {(1-u_{\beta_B}^2)} \partial_{u_{\beta_B}}
+ \frac{1} {1-u_{\beta_B}^2} (\partial_{\alpha}^2 +
\partial_{\gamma_B}^2 - 2 u_{\beta_B} \partial_{\alpha} \partial_{\gamma_B}))
(\frac{1}{2 \mu_R R^2}+\frac{1}{2 \mu_{2B} R_{2B}^2}) \nonumber \\
& - &  \hbar^2  (\partial_{u_{\theta{1A}}} {(1-u_{\theta_{1A}}^2)}
\partial_{u_{\theta{1A}}} + \frac{1} {1-u_{\theta_{1A}}} \partial_{\gamma_A}^2)
(\frac{1} {2 \mu_{1A} R_{1A}^2} +
\frac{1} {2 \mu_{2A} R_{2A}^2}) \nonumber \\
& - &  \hbar^2  (\partial_{u_{\theta{1B}}} {(1-u_{\theta_{1B}}^2)}
\partial_{u_{\theta{1B}}} + \frac{1} {1-u_{\theta_{1B}}} \partial_{\gamma_B}^2)
(\frac{1} {2 \mu_{1B} R_{1B}^2} +
\frac{1} {2 \mu_{2B} R_{2B}^2}) \nonumber \\
\end{eqnarray}

\begin{eqnarray}
\hat{T}_2 = 
&  & \frac{ \hbar^2} {2 \mu_{2A} R_{2A}^2} 
[- (\partial_{\alpha} + (x \partial_y -y \partial_x))
\frac{u_{\theta_{1A}}}{\sin \theta_{1A}}
\frac{\cos \gamma_A}{\sin \beta_A} \partial_{\gamma_A} \nonumber \\
 & +  & \partial_{u_{\beta_A}} \frac{u_{\theta_{1A}}}{\sin \theta_{1A}}
 \sin \beta_A  \sin \gamma_A \partial_{\gamma_A}
- 2 \partial_{\gamma_A} \frac{u_{\beta_A}}{\sin \beta_A}
\cos \gamma_A \frac{u_{\theta_{1A}}}{\sin \theta_{1A}} \partial_{\gamma_A}
\nonumber \\
 & + & 2 \partial_{\gamma_A}^2  + (\partial_{\alpha} + (x \partial_y
 -y \partial_x))
\frac{\sin \gamma_A}{\sin \beta_A} \sin \theta_{1A} \partial_{u_{\theta_{1A}}} 
\nonumber \\
& + & \partial_{u_{\beta_A}} \sin \beta_A \cos \gamma_A
\sin \theta_{1A} \partial_{u_{\theta_{1A}}}
+ \partial_{\gamma_A} \frac{u_{\beta_A}}{\sin \beta_A}
\sin \gamma_A \sin \theta_{1A} \partial_{u_{\theta_{1A}}} \nonumber \\
& - & \partial_{\gamma_A} \frac{u_{\theta_{1A}}}{\sin \theta_{1A}}
\frac{\cos \gamma_A}{\sin \beta_A}
(\partial_{\alpha} + (x \partial_y - y \partial_x)) \nonumber \\
 & + & \partial_{\gamma_A} \frac{u_{\theta_{1A}}}{\sin \theta_{1A}}
\sin \beta_A \sin \gamma_A \partial_{u_{\beta_A}} \\
 & + & \partial_{u_{\theta_{1A}}}
\frac{\sin \gamma_A}{\sin \beta_A} \sin \theta_{1a}
(\partial_{\alpha}  + (x \partial_y -y \partial_x)) \nonumber \\
& + & \partial_{u_{\theta_{1a}}}
\sin \beta_A \cos \gamma_A \sin \theta_{1A}
\partial_{u_{\beta_A}} + \partial_{u_{\theta_{1A}}}
\frac{u_{\beta_A}}{\sin \beta_A}
\sin \gamma_A \sin \theta_{1A} \partial_{\gamma_A}] \nonumber \\
& + &  \frac{\hbar^2} { 2 \mu_{2B} R_{2B}^2}
[\partial_{\alpha} \frac{u_{\theta_{1B}}}{\sin \theta_{1B}}
 \frac{\cos \gamma_B}{\sin \beta_B}
\partial_{\gamma_B}
+ \partial_{u_{\beta_B}} \frac{u_{\theta_{1B}}}{\sin \theta_{1B}}
 \sin \beta_B  \sin \gamma_B \partial_{\gamma_B}
\nonumber \\
& - &  2 \partial_{\gamma_B} \frac{u_{\theta_{1B}}}{\sin \theta_{1B}}
\cos \gamma_B \frac{u_{\beta_B}}{\sin \beta_B}
\partial_{\gamma_B} \nonumber \\
 & + & 2 \partial_{\gamma_B}^2  - \partial_{\alpha}
\frac{\sin \gamma_B}{\sin \beta_B} \sin \theta_{1B}
\partial_{u_{\theta_{1B}}} + \partial_{u_{\beta_B}} \sin \beta_B \cos \gamma_B
\sin \theta_{1B} \partial_{u_{\theta_{1B}}}
\nonumber \\
& + & \partial_{\gamma_B} \frac{u_{\beta_B}}{\sin \beta_B}
\sin \gamma_B \sin \theta_{1B} \partial_{u_{\theta_{1B}}} \nonumber \\
& + & \partial_{\gamma_B} \frac{u_{\theta_{1B}}}{\sin \theta_{1B}}
\frac{\cos \gamma_B}{\sin \beta_B} \partial_{\alpha}
+ \partial_{\gamma_B} \frac{u_{\theta_{1B}}}{\sin \theta_{1B}}
 \sin \beta_B \sin \gamma_B \partial_{u_{\beta_B}} \nonumber \\
& - & \partial_{u_{\theta_{1B}}}
\frac{\sin \gamma_B}{\sin \beta_B} \sin \theta_{1B}
\partial_{\alpha} + \partial_{u_{\theta_{1B}}}
\sin \beta_B \cos \gamma_B \sin \theta_{1B}
\partial_{u_{\beta_B}} \nonumber \\
& + & \partial_{u_{\theta_{1B}}}
\frac{u_{\beta_B}}{\sin \beta_B}
\sin \gamma_B \sin \theta_{1B}  \partial_{\gamma_B}] \nonumber \\
\end{eqnarray}

\begin{eqnarray}
 \hat{T}_3 &  = & \frac{\hbar^2}{2}  [ - 2\partial_{\alpha}
\cos{\alpha} \frac{u_{\beta_A}}{\sin \beta_A}
\frac{u_{\beta_B}}{\sin \beta_B} \partial_{\alpha} \nonumber \\
& - & (x \partial_y -y \partial_x)
\cos{\alpha} \frac{u_{\beta_A}}{\sin \beta_A}
\frac{u_{\beta_B}}{\sin \beta_B} \partial_{\alpha} \nonumber \\
& - & \partial_{\alpha}
\cos{\alpha} \frac{u_{\beta_A}}{\sin \beta_A}
\frac{u_{\beta_B}}{\sin \beta_B} (x \partial_y -y \partial_x) \nonumber \\
& + & 2 \partial_{\gamma_A} \frac{\cos{\alpha}}{\sin \beta_A \sin \beta_B}
\partial_{\gamma_B}
 +  \partial_{\alpha} \sin{\alpha} \frac{u_{\beta_A}}{\sin \beta_A}
\sin \beta_B \partial_{u_{\beta_B}} \nonumber \\
& + & \partial_{u_{\beta_B}} \sin{\alpha} \frac{u_{\beta_A}}{\sin \beta_A}
\sin \beta_B \partial_{\alpha}
+ (x \partial_y -y \partial_x)
\sin{\alpha} \frac{u_{\beta_A}}{\sin \beta_A}
\sin \beta_B \partial_{u_{\beta_B}} \nonumber \\
& + & \partial_{u_{\beta_B}} \sin{\alpha} \frac{u_{\beta_A}}{\sin \beta_A}
\sin \beta_B (x \partial_y - y \partial_x)
\nonumber \\
& + & \partial_{\alpha} \cos{\alpha} \frac{u_{\beta_A}}{\sin \beta_A}
\frac{1}{\sin \beta_B} \partial_{\gamma_B}
+ \partial_{\gamma_B} \cos{\alpha} \frac{u_{\beta_A}}{\sin \beta_A}
\frac{1}{\sin \beta_B} \partial_{\alpha} \nonumber \\
 & + &  \partial_{u_{\beta_A}} \sin{\alpha} \sin \beta_A
\frac{u_{\beta_B}}{\sin \beta_B} \partial_{\alpha}
+ \partial_{\alpha} \sin{\alpha} \sin \beta_A
\frac{u_{\beta_B}}{\sin \beta_B} \partial_{u_{\beta_A}}
 \nonumber \\
& + & 2 (x \partial_y -y \partial_x) \cos{\alpha} \frac{u_{\beta_A}}{\sin \beta_A}
\frac{1}{\sin \beta_B} \partial_{\gamma_B} \nonumber \\
& + & \partial_{u_{\beta_A}} \cos{\alpha} \sin \beta_A
\sin \beta_B \partial_{u_{\beta_B}}
+ \partial_{u_{\beta_B}} \cos{\alpha} \sin \beta_A
\sin \beta_B  \partial_{u_{\beta_A}} \nonumber \\
 & - &  \partial_{u_{\beta_A}} \sin{\alpha}
\frac{\sin \beta_A}{\sin \beta_B} \partial_{\gamma_B}
- \partial_{\gamma_B} \sin{\alpha}
\frac{\sin \beta_A}{\sin \beta_B} \partial_{u_{\beta_A}} \nonumber \\
& - & \partial_{\gamma_A} \cos{\alpha} \frac{1}{\sin \beta_A}
\frac{u_{\beta_B}}{\sin \beta_B} \partial_{\alpha}
- \partial_{\alpha} \cos{\alpha} \frac{1}{\sin \beta_A}
\frac{u_{\beta_B}}{\sin \beta_B} \partial_{\gamma_A}  \nonumber \\
& + &   \partial_{\gamma_A} \sin{\alpha}
\frac{\sin \beta_B}{\sin \beta_A} \partial_{u_{\beta_B}}
+  \partial_{u_{\beta_B}} \sin{\alpha}
\frac{\sin \beta_B}{\sin \beta_A}  \partial_{\gamma_A} \nonumber \\
 & - & 2 \partial^2_{\alpha}
- 2 \partial_{\alpha} (x \partial_y -y \partial_x) ] \nonumber \\
\end{eqnarray}

\begin{eqnarray}
\hat{T}_4 & = &  \frac{1} {2 \mu_R R^2}
(y^2 \partial_z^2 + z^2 \partial_y^2 + z^2 \partial_x^2
 + x^2 \partial_z^2 + x^2 \partial_y^2 + y^2 \partial_x^2 \nonumber \\
 & & -y \partial_y \partial_z z - \partial_y y z \partial_z 
- \partial_x x z \partial_z -x \partial_x  \partial_z z - 
x \partial_x \partial_y y - \partial_x x y \partial_y)
\nonumber \\
  & + &    \frac{1}{2 \mu_R R^2} [  (-
2 \frac{u_{\beta A}}{\sin \beta_A} { \partial_{\alpha}}
 - 2 \frac{1}{\sin \beta_A}  {\partial_{{\gamma}_A}}
\nonumber \\
& + & \frac{u_{\beta B}}{\sin \beta_B} \cos \alpha  { \partial_{\alpha}}
+{ \partial_{\alpha}} \frac{u_{\beta B}}{\sin \beta_B} \cos \alpha
- \sin \alpha \sin \beta_B { \partial_{u_{\beta B}}}
-  { \partial_{u_{\beta B}}} \sin \alpha \sin \beta_B
 - 2 \frac{\cos \alpha}{\sin \beta_B}  {\partial_{{\gamma}_B}}
) \nonumber \\
 & & (y  
{ \partial_z} - z { \partial_y}) \nonumber \\
& + & ( \sin \beta_A  { \partial_{u_{\beta A}}}
+  { \partial_{u_{\beta A}}}  \sin \beta_A
\nonumber \\
& + & \sin \alpha \frac{u_{\beta B}}{\sin \beta_B} 
 { \partial_{\alpha}}
+  { \partial_{\alpha}} \sin \alpha
 \frac{u_{\beta B}}{\sin \beta_B}
+ \cos \alpha \sin \beta_B  {\partial_{u_{\beta B}}}
+ { \partial_{u_{\beta B}}} \cos \alpha \sin \beta_B
- 2 \frac{\sin \alpha} {\sin \beta_B}
  { \partial_{{\gamma}_B}}) \nonumber \\
& & (z  { \partial_x} - x 
{ \partial_z} )\nonumber \\
 & + & 2 (x^2 \partial_y^2 + y^2 \partial_x^2 
- x \partial_x \partial_y y - \partial_x x y \partial_y)] \nonumber \\
&-& \frac{1}{2 \mu_R R^2} [ \frac{u_{\beta A}}{\sin \beta_A} 
(x \partial_y y \partial_z + x y \partial_y \partial_z
 -2 x \partial_y^2 z -2 \partial_x y^2 \partial_z + 
\partial_x y \partial_y z + \partial_x \partial_y y z)]
\nonumber \\
\end{eqnarray}

This completes the derivation of the kinetic energy operator
of H$_5$O$_2^+$. We emphasize again that this operator is exact.
Furthermore, the correctness of the derivation 
and implementation of the KEO was checked by comparing it
with numerical results provided by the program TNUM \cite{lau02:8560}.
A KEO can formally be written as 
   \mbox{
       $\hat{T}= (1/2)\sum_{ij}  G_{ij}(\mathbf{q})\partial_i\partial_j
       + \sum_j F_j(\mathbf{q})\partial_j
       +  V_{extra}(\mathbf{q})$}.
          TNUM computes $G$, $F$ and $V_{extra}$ numerically.
          We have checked that the numerical
          values of all the functions $G_{ij}(\mathbf{q})$ at several
          non-symmetrical grid points $\mathbf{q}$ agree with those provided
          by the program TNUM. The functions $F_j(\mathbf{q})$ are determined
          through the hermiticity of the KEO. As our KEO is obviusly hermitian,
          there is no need to check the $F_j$.

  \subsection{Hierarchical Representation of the Potential using Mode-Combination} \label{sec:cluster}

  The exact PES ($v(\mathbf{q})$) for {\zun} is a function of the 15 internal
  coordinates previously defined, where ``exact" refers to the full
  dimensional PES of Bowman and collaborators \cite{hua05:044308}.
  The calculation of vibrational levels or the IR spectrum
  requires a high accuracy in the representation of the
  Hamiltonian. The exact and trivial representation of $v(\mathbf{q})$ on
  a product grid is unfortunately beyond current
  computational capabilities: in this case the potential would be given on a
  grid of $\approx10^{15}$ points ($\approx10^{4}$ TB of disk space). 
  A direct use of the potential is hence impossible and in particular
  it is impossible to
  convert the potential to MCTDH product form by using the potfit
  algorithm \cite{jae96:7974,jae98:3772} since the full product grid is needed for such a 
  transformation.
  The problem of representing a high-dimensional PES
  for quantum-dynamical computation has already been considered in the context
  of Multimode simulations \cite{bow03:533}, as well as in the more general
  context of the high dimensional model representation (HDMR)
  \cite{rab99:197,ali01:127,li06:2474,man06:084109}.
  
  In such hierarchical representations a multidimensional function dependent on
  $f$ variables is approximated as:
  \begin{equation}
     \label{eq:hdmr}
     \tilde{v}(\mathbf{q}) = v^{(0)} + \sum_{\alpha=1}^f v^{(1)}_{\alpha}(q_\alpha) + 
                           \sum_{\alpha<\beta}^f v^{(2)}_{\alpha\beta}(q_\alpha,q_\beta)
     + \sum_{\alpha<\beta<\gamma}^f v^{(3)}_{\alpha\beta\gamma}(q_\alpha,q_\beta,q_\gamma) \cdots
  \end{equation}
  where $\mathbf{q}$ is a coordinate vector and $\tilde{v}(\mathbf{q})$ denotes the
  hierarchical approximation to $v(\mathbf{q})$.
  The component functions in Eq. (\ref{eq:hdmr}) can be determined by 
  minimizing the functional \cite{ali01:127,man06:084109}
  \begin{equation}
     \label{eq:hdmrErr}
     \int_D \left[ v(\mathbf{q}) - \tilde{v}(\mathbf{q}) \right]^2 w(\mathbf{q})d\mathbf{q},
  \end{equation}
  where $w(\mathbf{q})$ is some weight function on the integration domain that determines the
  form of the component functions. Taking $w(\mathbf{q})=\prod_{\kappa}^f w_\kappa(q_\kappa)$
  leads to component functions of the form \cite{ali01:127,wan03:4707,li06:2474,man06:084109}
  \begin{subequations}
     \label{eq:hdmrComp}
  \begin{eqnarray}
     v^{(0)}  & = & 
                \int_D \prod_{\kappa=1}^f w_\kappa(q_\kappa) v(\mathbf{q})\,d\mathbf{q} \\
     v^{(1)}_{\alpha}(q_\alpha) & = & 
                \int_{D^{f-1}} \prod_{\kappa\neq\alpha}^f w_\kappa(q_\kappa) 
                       v(\mathbf{q})\,d\mathbf{q}^{\alpha} - v^{(0)} \\
     v^{(2)}_{\alpha\beta}(q_\alpha,q_\beta) & = & 
                \int_{D^{f-2}} \prod_{\kappa\neq\alpha,\beta}^f w_\kappa(q_\kappa) 
               v(\mathbf{q})\,d\mathbf{q}^{\alpha\beta}
             - v^{(1)}_{\alpha}(q_\alpha) - v^{(1)}_{\alpha}(q_\alpha) - v^{(0)}\\
  \nonumber \cdots & & 
  \end{eqnarray}
  \end{subequations}
  There $\mathbf{q}^{\alpha}$ represents a vector of coordinates in which the $\alpha$-th component
  has been removed, $\mathbf{q}^{\alpha\beta}$ represents a vector in which $\alpha$ and $\beta$ components
  have been removed, and so on.
  Non-separable weights have also been considered, which lead to more complicated expressions for 
  component functions (see for example the Appendix in Ref. \onlinecite{man06:084109}).
  It is possible to evaluate the integrals in Eq. (\ref{eq:hdmrComp}) by random sampling
  of the coordinate space, leading to the so called RS-HDMR \cite{ali01:127,wan03:4707,li06:2474,man06:084109}.
  However, the cumbersome multi-dimensional integrals can be solved trivially
  by choosing the particular weight
  \begin{equation}
     \label{eq:hdmrCutDelta}
     w(\mathbf{q}) = \prod_{\alpha=1}^f \delta(q_\alpha-a_\alpha)
  \end{equation}
  which corresponds to the cut-HDMR approximation \cite{bow03:533,rab99:197,li01:1,ali01:127}.
  There $a_\alpha$ is the $\alpha$-th component of point $\mathbf{a}$, the reference expansion point
  in coordinate space. 
  Using the definition of $w(\mathbf{q})$ in Eq. (\ref{eq:hdmrCutDelta}), the different 
  terms in Eq. (\ref{eq:hdmr}), which we will also 
  refer to as uncombined clusters (UC), are given up to second order by \cite{li01:1}
  \begin{subequations}
     \label{eq:hdmr-terms}
  \begin{eqnarray}
               v^{(0)} &=& v(\mathbf{a}) \\
               v^{(1)}_\alpha(q_\alpha) &=& v(q_\alpha;\mathbf{a}^\alpha) - v^{(0)} \\
               v^{(2)}_{\alpha\beta}(q_\alpha,q_\beta) 
                    &=& v(q_\alpha,q_\beta;\mathbf{a}^{\alpha\beta}) 
                     - v^{(1)}_\alpha(q_\alpha) - v^{(1)}_\beta(q_\beta) - v^{(0)},
  \end{eqnarray}
  \end{subequations}
  where higher orders follow trivially.
  Cut-HDMR representations of a PES -- also called n-mode representation -- have been already used 
  successfully to accurately compute vibrational
  energy-levels of molecular systems \cite{bow03:533,cha04:2071} and reaction rates
  for molecule-surface scattering \cite{kro07:334}.
  Unfortunately, the use of a HDMR representation of the PES of the form of Eq. (\ref{eq:hdmr}) leads to
  a combinatorial increase in the number of terms as the order of correlation increases. At the
  same time, higher order terms are given on grids with an exponentially increasing number of points,
  which leads to quick stagnation in the maximum correlation order that can be practically included.

  Instead of directly adopting the expression in Eq. (\ref{eq:hdmr}) for $v(\mathbf{q})$
  we make use of the fact that the wavefunction in Eq. (\ref{eq:ansatz}) is given
  in terms of combined modes. Hence we define the potential also as a function of
  the combined modes.
  The use of combined modes instead of coordinates as base of the hierarchical expansion
  attenuates the combinatorial increase in complexity
  found when using Eq. (\ref{eq:hdmr}) while still retaining the simple evaluation 
  of the expansion terms   given by Eq. (\ref{eq:hdmr-terms}) and the inclusion 
  of high-order correlations. 
  
  For $f$ coordinates $\mathbf{q}=[q_1,\ldots, q_f]$, $p$ particles or combined modes 
  $\mathbf{Q}=[Q_1,\ldots, Q_p]$ are defined, such that
  $Q_i = [q_1^{(i)},\ldots,q_{f_i}^{(i)}]$ and $\sum_{i=1}^p f_i = f$.
  The reference potential can be equivalently given as a function of the
  combined modes, $V(\mathbf{Q}) \equiv v(\mathbf{q})$.
  The general hierarchical expansion of Eq. (\ref{eq:hdmr}) is now written in
  terms of the combined modes $Q_i$, instead of coordinates $q_\alpha$,
  which up to second order reads:
  \begin{equation}
    \label{eq:hdmr-VQ}
    \tilde{V}(\mathbf{Q}) = V_0 + \sum_{i}  V_i^{(1)}(Q_i)
                              + \sum_{ij} V_{ij}^{(2)}(Q_i,Q_j).
  \end{equation}
  First order $V_i^{(1)}$ and second order $V_{ij}^{(2)}$ combined clusters (CC)
  are defined in an analogous way to Eq. (\ref{eq:hdmr-terms}):
  \begin{subequations}
     \label{eq:VQ-terms}
  \begin{eqnarray}
               V^{(0)} &=& V(\mathbf{a}) \\
               V_i^{(1)}(Q_i) &=& V(Q_i;\mathbf{a}^i) - V_0   \\
               V_{ij}^{(2)}(Q_i,Q_j) &=& 
                      V(Q_i,Q_j;\mathbf{a}^{ij})
                         - V_i^{(1)}(Q_i) - V_j^{(1)}(Q_j) - V_0
  \end{eqnarray}
  \end{subequations}
  First order CC can be given directly in multidimensional product-grids
  which are direct products of the corresponding 1D DVR grids since the wavepacket
  is expanded in sums of products of SPFs which are defined on the same combined grids as
  used in the cluster expansion.
  Second order and higher CC are conveniently and accurately represented as sums of products
  of multidimensional mode-functions by employing the potfit-algorithm \cite{jae96:7974,jae98:3772}.
  
\begin{figure}[h!]
  \begin{center}
    \includegraphics[width=8.5cm]{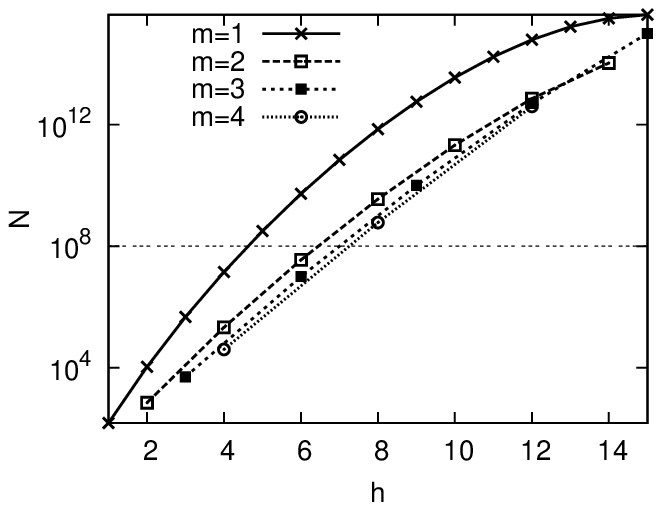}
  \end{center}
 \caption{\figfoot}
 \label{fig:mode-comb}
\end{figure}

  On the other hand, first order CC can be exactly represented 
  as a cut-HDMR expansion of the form
  of Eq. (\ref{eq:hdmr}) up to order $f_i$ in terms of the 
  UC made of the coordinates $[q_1^{(i)},q_2^{(i)},\ldots]$
  in $Q_i$, while analogously, 
  second order CC can be given by a cut-HDMR expansion up to order 
  $f_i+f_j$ in terms of the corresponding UC.
  When inspecting the expansion in Eq. (\ref{eq:hdmr-VQ}) one realizes that
  the correlation between three coordinates belonging to different modes, 
  i.e. $[q_\alpha^{(i)},q_\beta^{(j)},q_\gamma^{(k)}]$, is accounted for up to second
  order with respect to the uncombined representation since only the UC
  $v^{(2)}_{\alpha\beta}(q_\alpha^{(i)},q_\beta^{(j)})$,
  $v^{(2)}_{\alpha\gamma}(q_\alpha^{(i)},q_\gamma^{(k)})$ and
  $v^{(2)}_{\beta\gamma}(q_\beta^{(j)},q_\gamma^{(k)})$ 
  are contained in the second-order CC
  $V_{ij}^{(2)}(Q_i,Q_j)$,
  $V_{ik}^{(2)}(Q_i,Q_k)$ and
  $V_{jk}^{(2)}(Q_j,Q_k)$ respectively.
  Thus, using a second order expansion of CC one implicitely introduces a
  cluster-selection scheme of higher (up to $f_i+f_j$) order 
  UC. The higher-order cluster-selection scheme is determined
  by how coordinates are grouped together into combined modes.
  The same reasoning can be extended to higher than second order CC.
  By combining coordinates which are strongly coupled it is then possible to get a 
  high-accuracy potential while still using a reduced number of CC.
    \placefig{fig:mode-comb}
  Assuming that the grid representation of each coordinate uses on
  average $N$ points, the number of points in coordinate-space, $N_{tot}$, used to define 
  the UC, i.e. without mode combination, is given by
  \begin{equation}
     \label{eq:mode-comb}
     N_{tot} = \sum_{\alpha=0}^{h} \binom{f}{\alpha} N^{\alpha},
  \end{equation}
  where $h$ is the maximum allowed order for the clusters and $f$ is the
  number of degrees of freedom.
  In case the coordinates are combined into modes with $m$ coordinates each, the
  number of grid points needed to represent the clusters is given by
  \begin{equation}
     \label{eq:mode-comb2}
     N_{tot} = \sum_{i=0}^{h/m} \binom{f/m}{i} N^{\,i\cdot m}.
  \end{equation}
  The number of points $N$ for different values of the parameters defining
  different clustering schemes is depicted in Fig. \ref{fig:mode-comb}.
  The horizontal axis represents $h$, the maximum order of clustering in terms
  of the uncombined coordinates, i.e., the maximum order of the UC in the expansion.
  A total of $f=15$ coordinates is assumed.
  A horizontal line is drawn at $10^8$, which, tentatively, constitutes a practical
  limit to the total number of grid points that can be used both concerning the
  generation of the clusters and their subsequent use in the dynamical calculations.
  The example in Fig. \ref{fig:mode-comb} assumes that there are $N=10$ points per coordinate.
  Using 3D modes it is possible to include 6th order UC with around $10^7$ grid points, which is 
  of the order of the PES presented in section \ref{sec:PES}. The inclusion of up
  to 5th order UC without mode-combination would require more than $10^8$
  points while the inclusion of 6th order UC would require around
  $10^{10}$ points. 
  We emphasize again, however, that the representations up
  to order $h$ with and without mode-combination are not equivalent. The representation
  without mode-combination contains {\em all} the possible UC of
  coordinates up to order $h$. In the case of using CC one is implicitly
  selecting a subset of UC.
  A definition of meaningful mode combinations should however be possible in most cases.
  Such a definition would be based
  on chemical common-sense, e.g., coordinates belonging to the same chemical group
  or to the same molecule in a cluster are good candidates to be combined.
  As a final remark,
  we note that there is ongoing
  effort by other groups \cite{li06:2474,man06:084109} to use
  parametrized functions instead of grids to describe the clusters.
  This approach may turn out to be more efficient than a grid representation.

  \subsection{Potential Energy Surface for the {\zun} cation} \label{sec:PES}
  
  In order to construct the PES for the {\zun} cation employing the approach described
  above one must start by defining the combined modes that are going to be used.
  In the present case the following five multidimensional modes are selected: 
  $Q_1 = [z,\alpha,x,y]$,
  $Q_2 = [\gamma_A,\gamma_B]$,
  $Q_3 = [R, u_{\beta_{A}}, u_{\beta_{B}}]$,
  $Q_4 = [R_{1 A}, R_{2 A}, u_{\theta_{1 A}}]$ and
  $Q_5 = [R_{1 B}, R_{2 B}, u_{\theta_{1 B}}]$.
  It is convenient that coordinates $x$, $y$ and $\alpha$ are grouped together due
  to symmetry conserving reasons which are exposed below.
  Modes $Q_2$ and $Q_3$ contain the wagging and rocking coordinates, respectively.
  Modes $Q_4$ and $Q_5$ contain the Jacobi coordinates which represent the internal configuration of
  each water molecule. 
  Coordinates $z$ and $R$ are good candidates to be combined together, as will be discussed in
  Section \ref{sec:quality}. They are not combined here since this would require the definition
  of a 6th mode. We have, in fact, started to do some tests with a 6 particle mode-combination, and we give
  some brief remark on this below.
  The definition of the underlying 1D grids is provided in Table \ref{tab:1Dgrids}.
  \placetab{tab:1Dgrids}
  
  Following the procedure outlined above one may select a reference point in coordinate
  space and proceed straightforwardly to the computation of the clusters defining the cluster expansion.
  Instead of this, the PES expansion is defined in terms of $M=10$ reference points
  and the weight in Eq. (\ref{eq:hdmrErr}) takes the form
  \begin{equation}
    \label{eq:weight}
    w(\mathbf{q}) = \frac{1}{M} \sum_{l=1}^M \delta(\mathbf{q}-\mathbf{a}_l).
  \end{equation}
  The reference points $\mathbf{a}_l$ are located on or very
  close to stationary points in the lowest energy regions of the PES.
  After Eq. (\ref{eq:weight}) the PES expansion is given by
  \begin{equation}
    \label{eq:VtotQ}
    \tilde{V}_{tot}(\mathbf{Q}) = \frac{1}{M} \sum_{l=1}^{M} \tilde{V}_l({\mathbf{Q}}).
  \end{equation}
  The $\tilde{V}_l({\mathbf{Q}})$ terms are given by Eqs. (\ref{eq:hdmr-VQ}) and (\ref{eq:VQ-terms}).
  The specific form of $\tilde{V}_l({\mathbf{Q}})$ that has been used here is given by
  \begin{equation}
    \label{eq:VtotQ2}
    \tilde{V}_l({\mathbf{Q}}) =  V_l^{(0)} 
                              +   \sum_{i=1}^5 V_{l,i}^{(1)}(Q_i) 
                              +   \sum_{i=1}^4 \sum_{j=i+1}^5  V_{l,ij}^{(2)}(Q_i,Q_j)
                              +   V_{l,z23}^{(3)}(z,Q_2,Q_3),
  \end{equation}
  where the modes $Q_1\cdots Q_5$ have been defined above.
  The $V_l^{(0)}$ term is the energy at the reference geometry $l$.
  The $V_{l,i}^{(1)}$ terms are the intra-group potentials obtained by
  keeping the coordinates 
  in other groups at the reference geometry $l$, while the $V_{l,ij}^{(2)}$
  terms account for the group-group correlations. The potential with up 
  to second-order terms gives already a very reasonable description of
  the system. The $V_{l,z23}^{(3)}$ term accounts for three-mode
  correlations between the displacement of the central proton, the
  distance between both water molecules and the angular wagging and
  rocking motions.
  Note that the primitive grids in each coordinate are the same irrespective of the reference point
  used to expand the potential. This means that the average, Eq. (\ref{eq:VtotQ}), can be carried out
  before the dynamical calculations by summing over all the generated grids of the same coordinates
  for each reference geometry, involving no extra cost for the dynamics.
  The justification for the multi-reference approach lies on the nature of {\zun}.
  The protonated water-dimer is a very floppy system featuring several equivalent minima 
  and large amplitude motions that traverse low potential energy barriers.
  Thus, the amount of configurational space available to the system at low vibrational energies
  is already large and then it is not well covered by a single reference point.
  The property that, for a single reference point, the PES expansion is exact at the 
  reference point and hypersurfaces involving
  the displacement of up to $h_m$ modes is lost after averaging over several
  reference geometries. However, the overall mean error is reduced by the averaging.
  
\begin{figure}[h!]
    \begin{center}
      \includegraphics[width=8.5cm]{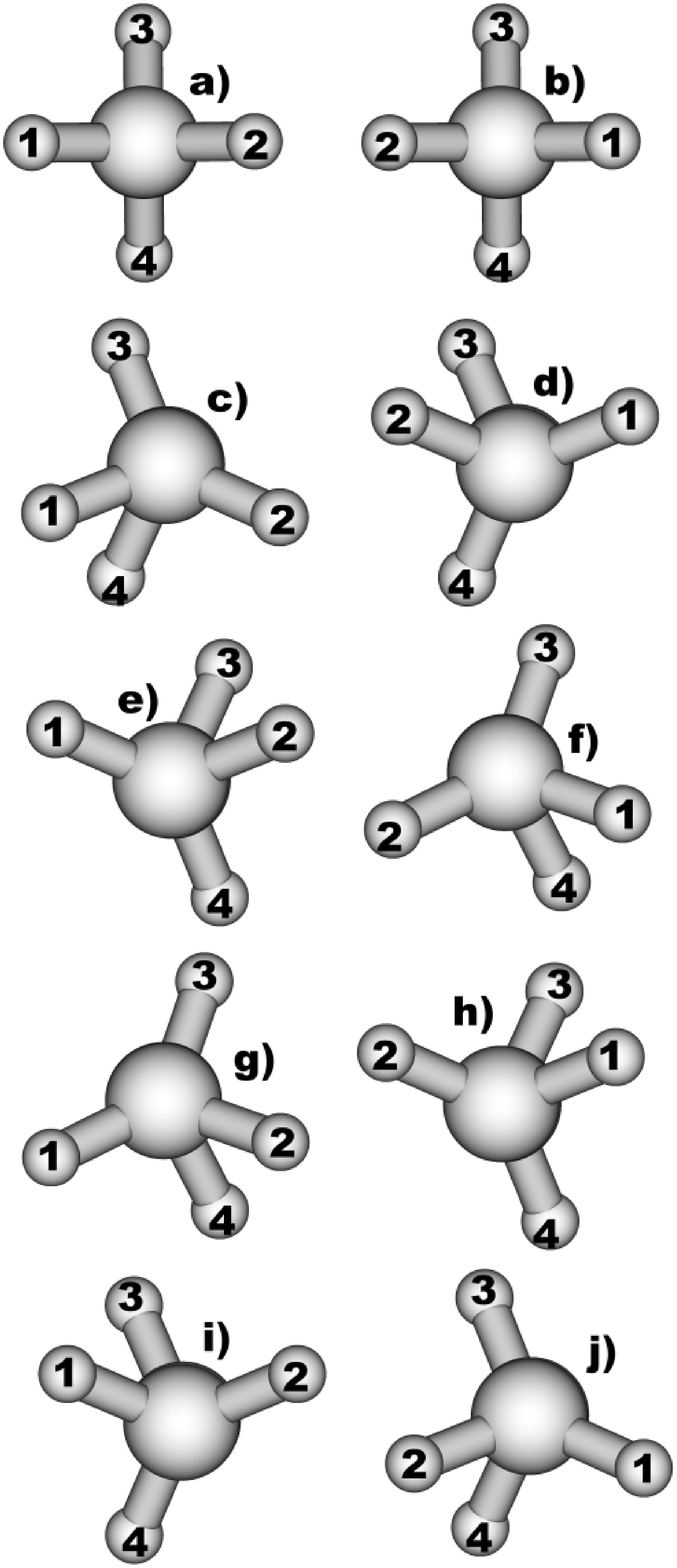}
    \end{center}
   \caption{\figfoot}
   \label{fig:refGeos}
\end{figure}

  The use of several reference points has a further implication which is 
  related to the symmetry properties of the system Hamiltonian.
  In the case of {\zun} a possible single reference geometry to define the PES expansion
  would be one of the eight equivalent absolute minima which belong to the ${\mathcal C}_{2}$
  point groups.
  This choice, however, breaks the total symmetry of the Hamiltonian. We have seen that
  cut-HDMR is exact at the reference point and all hypersurfaces
  in which up to $h_m$ modes have been displaced from the reference point, where $h_m$ is 
  the expansion order in terms of CC.
  The description of the rest of ${\mathcal C}_{2}$ points when one is 
  used as a reference is thus not equivalent 
  and the whole symmetry is broken.
  A possible solution would be to use as reference one of the
  two equivalent ${\mathcal D}_{2d}$ stationary points, which lie around 300 {\icm} above the 
  absolute ${\mathcal C}_{2}$ minima.
  They correspond to $\alpha=90$ or $270$ degrees, respectively.
  Using only one of the ${\mathcal D}_{2d}$ points as reference results 
  in a similar breakage of the total symmetry
  of the Hamiltonian. The same happens again if one of the two 
  equivalent ${\mathcal D}_{2h}$ ($\alpha=0$ or $180$ deg.) stationary points is
  used, which are even higher in energy.
  One should note that this is a highly symmetrical, multiminima system 
  in which several permutations of identical
  particles are possible through crossing of low energy conformational barriers.
  The vibrational levels of {\zun} can
  be labeled according to the permutation-inversion symmetry
  group ${\mathcal G}_{16}$ \cite{bunker-book:symmetry,wal99:10403}, 
  which contains the ${\mathcal D}_{2d}$ point group as a subgroup,
  but additionally allows to permute the H-atoms of either of the
  two monomers.
  The two ${\mathcal D}_{2d}$ and eight equivalent ${\mathcal C}_{2}$ geometries
  which are used as reference points are depicted
  in Fig. \ref{fig:refGeos}. The difference between the structures at the left and right columns
  is a 180 degrees rotation of $\alpha$, or equivalently the permutation of
  the two hydrogens of one of the water moieties.
  The way out of the symmetry breakage problem is to use all the structures depicted in 
  Fig. \ref{fig:refGeos} as reference points of the cluster
  expansion. The final potential is given then, as discussed above, by the average with respect
  to all the reference points. 
  By using this set of reference points the symmetry of the original PES is maintained. Indeed, it
  will be maintained as long as, for an arbitrary selected reference geometry, all the symmetry
  equivalent points generated by the permutations-inversions of the ${\mathcal G}_{16}$ group are also
  considered as reference points.
  \placefig{fig:refGeos}

\section{Results and Discussion}

  \subsection{Quantum Dynamical Calculations}  \label{sec:ZPE}

   The kinetic energy and potential energy operators already discussed are used
   to compute the zero point energy (ZPE) of the system and the corresponding ground-state 
   vibrational-wavefunction.
   The algorithm that implements the computation of eigenvalues and eigenfunctions of the
   system Hamiltonian within the MCTDH program is called {\em improved relaxation} and is
   described elsewhere \cite{mey06:179}. This algorithm is essentially a multiconfiguration 
   self-consistent field approach that takes advantage of the MCTDH machinery.
   All the reported simulations were performed with the Heidelberg MCTDH package of programs
   \cite{mctdh:package}.
   
   The comparison between the largest, converged MCTDH calculation and other reported
   results on the same PES is given in Table \ref{tab:zpe}. As a reference we take the
   given diffusion Monte Carlo (DMC) result \cite{mc05:061101} which has an associated 
   statistical uncertainty of 5 {\icm}. A simple normal-modes analysis (NMA) with
   normal modes constructed from the Hessian matrix taken at the
   $\mathcal{C}_{2}$ minimum yields a ZPE of $12\,635$ {\icm}, {\em only} $242$ {\icm} larger than the
   DMC result. We believe that this surprisingly good result arises from fortuitous error cancellation.
   As will be discussed below, 3 out of the 15 internal degrees of freedom 
   ($\gamma_a$, $\gamma_b$ and $\alpha$) are not described by a single-well in the lowest
   energy region of the potential, while the proton-transfer motion ($z$)
   features a nearly quartic potential which is strongly coupled to the water-water stretching ($R$).
   The most comprehensive calculations on the vibrational ground state based on a wavefunction
   approach to date are those of Bowman and collaborators \cite{mc05:061101} 
   using the Multimode program \cite{bow03:533}. The vibrational configuration interaction (VCI) 
   results, both using the single reference (SR) and reaction path (RP) variants are
   found in Table \ref{tab:zpe}. These calculations use a normal-mode based Hamiltonian. 
   They incorporate correlation between the different degrees of freedom due to the
   cluster expansion of the potential \cite{bow03:533} and the use of a CI wavefunction.
   The best reported VCI result for the ZPE lies still 104 {\icm} above the DMC value.
   It is worth to mention that before switching to a Hamiltonian based on polyspherical coordinates
   we tried a Hamiltonian expressed in rectilinear-coordinates and obtained
   results similar to those of Bowman and collaborators.
   \placetab{tab:zpe}
   The MCTDH converged result for the ZPE is given in Table \ref{tab:zpe}.
   The obtained value for the ZPE is $12\,376.3$ {\icm}, $16.7$ {\icm} below
   the DMC value. Table \ref{tab:zpe-mctdh} contains ZPE values obtained using an
   increasing number of configurations. According to these results the MCTDH
   reported values are assumed to be fully converged with respect to the number of configurations.
   The deviation from the DMC result must be attributed to the cluster expansion
   of the potential, Eqs. (\ref{eq:VtotQ},\ref{eq:VtotQ2}).
   \placetab{tab:zpe-mctdh}

We give here some technical details regarding the largest
MCTDH calculation with $10\,500\,000$ configurations, which is probably
one of the largest MCTDH calculations performed to date in terms of required
computational resources. The calculation was made using
the recent parallel version of the MCTDH code which is still under
development in our group. The calculation was run on a 8-processor,
shared-memory machine with Intel Itanium2 processors, and used a total 
amount of $13.5$ Gb of main memory.
4 Gb were devoted to storage of the 
Krylov vectors and Krylov times Hamiltonian  vectors
of the Davidson diagonalization procedure.
The rest was needed for the representation of the Hamiltonian and the mean fields, 
the vector of coefficients, SPFs and work arrays.
Each step of the wavefunction relaxation lasted around $13.5$ hours of
wall-clock time,
and consisted of the aforementioned Davidson diagonalization of the system
Hamiltonian in the current basis of SPFs followed by a 2 fs imaginary-time propagation
of the SPFs. The whole procedure was iterated until self-consistency of the expansion
coefficients and SPFs was reached.

In contrast, the smallest calculation reported in Table \ref{tab:zpe-mctdh}, which used
$172\,800$ configurations, consumed a total amount of memory of around $263$ Mb.
Each diagonalization of the system Hamiltonian plus imaginary time propagation of the
SPFs lasted $10$ minutes of wall clock time using two processors in parallel on the
same Intel Itanium2 cluster. The comparison of the two aforementioned simulations illustrates
one of the major strengths of MCTDH, namely, the usage of variationally optimal coefficients
and orbitals leads to an early convergence with respect to the size of the multiconfigurational
expansion. Relatively good results (only $7.4$ {\icm} energy difference 
between the largest and smallest calculation) are already obtained using very
moderate computational resources.

   \subsection{Properties of the ground state of the system}  \label{sec:wavefunction}
\begin{figure}[h]
  \begin{center}
    \includegraphics[width=12.75cm]{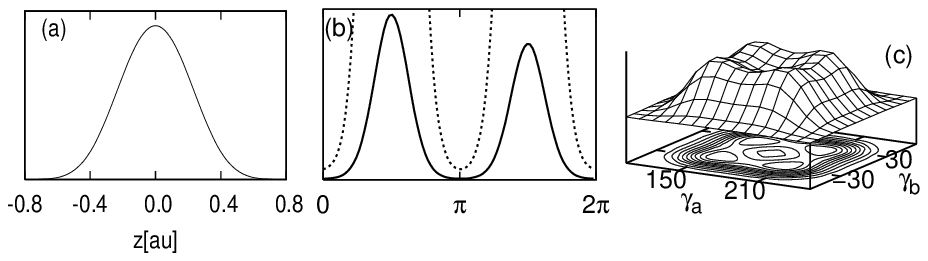}
  \end{center}
 \caption{\figfoot}
 \label{fig:wavefunc}
\end{figure}

   The probability density of the ground state wavefunction with respect to
   some selected coordinates and integration over the remaining coordinates is
   given in Fig. \ref{fig:wavefunc}. Fig. \ref{fig:wavefunc}a shows the density
   along the proton-transfer coordinate $z$. The probability density is non negligible in a
   range spanning about 1 bohr.
   Fig. \ref{fig:wavefunc}b depicts
   the density along the $\alpha$ internal rotation coordinate. Along this
   coordinate the system interconverts between two equivalent regions of configurational space.
   The barrier corresponds to planar configurations of the whole system and is
   about 300 {\icm} high depending on the configuration of
   the rest of coordinates.
   The system can interconvert between both halves even when in
   the ground vibrational state.
   The dotted curve in Fig. \ref{fig:wavefunc}b depicts the density at a 10 times enlarged scale and
   clearly shows a non vanishing density for $\alpha=0,\pi$. The splitting state arising from the
   barrier along $\alpha$ has also been computed and the splitting energy has been found to be 1 {\icm}.
   The small asymmetry observed in the density in Fig. \ref{fig:wavefunc}b has been analyzed. It arises
   from the fact that the position of the proton is defined relative to one of the two water monomers,
   monomer A, in order to have 15 internal coordinates (see Eq. (\ref{Eq:485}) and related text).
   Consequently, a change in $\alpha$ rotates the central proton so 
   that its relative position to monomer A remains
   unaltered. In contrast, in order for the relative position of 
   the central proton and monomer B to remain unaltered
   during a change in $\alpha$, coordinates $x$ and $y$ must change accordingly.
   We emphasize that the kinetic operator is still exact and not affected by this. 
   However, $\alpha$, $x$ and $y$
   coordinates are defined on non-matching grids. The total symmetry between monomers A and B
   with respect to the central proton is conserved in the case of a continuous configurational
   space instead of a discretized one. Thus, symmetry is slightly broken due to discretization
   of the configurational space. Energetically, the effect of this symmetry breakage must be well below
   1 {\icm} which is the splitting caused by the barrier along $\alpha$, since a larger perturbation
   would localize the density on one side of the barrier breaking the double well feature.
   A perfect symmetry after discretization of the coordinates would probably be
   obtained if one switches to cylindrical coordinates ($z$, $\rho$, $\phi$) for the
   proton and uses identical grids for the angles $\alpha$ and $\phi$.
   
   Fig. \ref{fig:wavefunc}c shows the probability density along the wagging coordinates. It consists of
   four equivalent maxima, each of them centered at around $\pm 30$
   degrees from the planar water configuration, so that both water molecules are in
   pyramidal configuration. The
   probability density corresponding to one of the two water molecules in a planar configuration
   is however quite large and indicates a high probability of exchange between equivalent
   configurations of the system in which the water molecules switch between pyramidal
   geometries. Each of these four density maxima corresponds roughly to one of the ${\mathcal C}_{2}$ 
   equivalent minima on the PES. A total of 8 equivalent ${\mathcal C}_{2}$ minima are present
   since the barrier along coordinate $\alpha$ divides the configurational space in two equivalent halves.
   When both monomers are in planar configuration the system interconverts between both 
   ${\mathcal D}_{2d}$ and ${\mathcal D}_{2h}$ configurations by rotation along $\alpha$.
    \placefig{fig:wavefunc}
    
   Only after the introduction of a Hamiltonian based on polyspherical coordinates 
   a satisfactory description the {\zun} cation was possible. 
   The use of polyspherical coordinates allows for
   the characterization of the system in terms of well defined stretching, bending, rocking, wagging and internal
   rotation motions, each of which corresponds to {\em a single} coordinate. This fact
   keeps the correlation between coordinates in the PES relatively small, and the relation between 
   different coordinates can be usually understood in simple physical terms. The representation of the
   WF given in these coordinates converges then more quickly than a WF constructed from rectilinear coordinates
   and can be more compact.
   The price to pay, however, is a much more complicated expression for the kinetic energy operator.


   \subsection{Quality of the PES expansion}  \label{sec:quality}
   In order to illustrate the convergence of the mode-combination based cluster-expansion
   PES introduced in Section \ref{sec:PES} the expectation values of the different
   terms of the potential are calculated for the ground vibrational state. These values 
   are given in {\icm} in Table \ref{tab:cluExpect}.
   The sum of the first order $\langle\Psi_0|V^{(1)}(Q_i)|\Psi_0\rangle$ terms is close to 6800 {\icm},
   half the ZPE, indicating that they carry the major weight in the description of the PES. The
   second order clusters introduce the missing correlation between modes.
   They have expectation values one order of magnitude smaller than the first order terms with
   one exception, the matrix element arising from the $V^{(2)}(Q_1,Q_2)$ potential. 
   This can be easily understood by
   noting that modes $Q_1$ and $Q_2$ contain coordinates $z$ and $R$, respectively.
   These two coordinates are strongly correlated and indeed they
   would be good candidates to be put in the same mode in an alternative mode-combination scheme.
   The only third order term that was introduced presents a rather marginal contribution to the
   potential energy of the system. These values prove that the PES representation used is of a good quality
   and rather well converged with respect to the reference PES, at least for the energy domain of interest.
   The square root of the expectation value of the potential squared is depicted in the third column.
   It is a measure of the dispersion around the expectation value and should also ideally vanish.
   The values indicate that the PES representation is good, albeit not yet fully
   converged. Some more terms of higher than second order may be added in the future as computational
   resources allow for it.
   \placetab{tab:cluExpect}

To finish the discussion it is worth mentioning that we have recently started to test
a new mode-combination. It is based on 6 modes instead of 5. The definition
of the modes in terms of the coordinates is as follows: 
  $\tilde{Q}_1 = [z,R]$,
  $\tilde{Q}_2 = [x,y,\alpha]$,
  $\tilde{Q}_3 = [\gamma_A,\gamma_B]$,
  $\tilde{Q}_4 = [u_{\beta_A}, u_{\beta_B}]$,
  $\tilde{Q}_5 = [R_{1 A}, R_{2 A}, u_{\theta_{1 A}}]$ and
  $\tilde{Q}_6 = [R_{1 B},R_{2 B}, u_{\theta_{1 B}}]$.
At the beginning we started with 5 modes in order to keep the number of configurations
as small as possible. 
This has the drawback that one of the combined modes, $Q_1=[z,\alpha,x,y]$ is very large, 
thus its SPFs are harder to
propagate. However, in the reported calculations we use the parallel version of the MCTDH code,
and the parts which can be parallelized most efficiently are those involved with the vector
of coefficients. Thus, a 6-modes scheme seems to be more efficient when the parallel code
is used. In the 6-modes scheme we include all the first, second and some
selected third order clusters. The obtained values of the ZPE are still preliminary but located in
the region of -10 {\icm} with respect to DMC, in good accordance to the results of the 5-mode scheme. 
Such results indicate indeed that the reported simulations are robust with respect to the mode-combination
scheme.

\section{Summary and Conclusion}  \label{sec:conclusions}

The protonated water-dimer ({\zun}) is studied in its full dimensionality (15D) 
by quantum-dynamical wavefunction-methods, using the MCTDH approach.
A set of curvilinear coordinates is used which accounts for the fluxional,
multi-minima nature of the cation.
An exact expression for the kinetic energy operator of the system is
derived using the polyspherical method. A discussion of the different steps
involved in the derivation is given.
The set of polyspherical coordinates introduced allows the description of the different
motions of the system, namely proton transfer, water-water stretching, OH stretchings, bendings and
internal rotation by a single coordinate each which also have clear geometrical meanings in terms
of angles or distances.
The PES used in the calculations is that of Huang et al., the most
accurate PES available for this system to date. The PES must be represented in a
numerically and computationally adequate way for the quantum-dynamical simulations on
a discrete multidimensional grid to be feasible. To this end, a variation of the hierarchical cut-HDMR method 
is presented which takes advantage of the mode-combination strategy used to represent 
the wavefunction. Combined modes, instead of the coordinates, are used to define
the hierarchical expansion. The expansion of the PES is seen to converge quickly with the
number of clusters in the expansion, which can be attributed to two factors:
\begin{itemize}
  \item[1.] A large amount of the correlation is already captured
            within the modes, which leads to quick convergence of the PES with respect to 
            the number of clusters in the expansion.
  \item[2.] The set of polyspherical coordinates used allows for the definition of physically
            meaningful modes avoiding at the same time artificial correlations
            (which are due to inadequate coordinates) to appear both in the potential energy and
            wave function.
\end{itemize}
The ZPE of the system is calculated and the results obtained
show an excellent agreement with previous DMC calculations on the same PES. 
The reported ZPE with MCTDH lies only $16.7$ {\icm} below the reported DMC result.
This deviation from the ``exact" DMC result (statistical error 5 {\icm}) must
be due to the potential representation, as the KEO is exact and as it is shown,
the reported ZPE is converged with respect to the number of configurations
in the MCTDH expansion. A fully converged MCTDH calculation needs a rather large amount
of computational resources, however, it is shown that reasonably good results are obtained with a much
smaller configurational space due to the optimality of both the coefficients and the SPFs
in the MCTDH expansion.
The properties of the ground vibrational state are analyzed. The central proton delocalizes on
a range of about 1 bohr along the water-water axis. The fluxional nature of the system is 
exemplified in the probability density along the wagging $\gamma_A$ and $\gamma_B$ coordinates. The 
2D space spanned by these
coordinates presents four density maxima which roughly correspond to the $\mathcal{C}_{2}$ minima
of the system. In going between different higher density regions the system switches between pyramidal
conformations of the water molecules. These conformational changes take place along
low potential-energy barriers. The system is completely delocalized over these low barriers 
leading to a highly symmetric ground-state wave function. 
The internal rotation coordinate $\alpha$ is seen to divide the configurational space
in two equivalent halves. The probability density for $\alpha=0$ and $\alpha=\pi$ (planar $\mathcal{D}_{2h}$)
is non-negligible for the ground vibrational state. The tunneling splitting arising from the barrier along
$\alpha$ is computed to be 1 {\icm}.
The convergence of the PES expansion is monitored with respect to the expectation values
of the potential-energy terms which define it, i.e. by inspecting 
$\langle \Psi_0 | V_{ij\cdots}| \Psi_0 \rangle$ and $\langle \Psi_0 | V_{ij\cdots}^2| \Psi_0 \rangle^{1/2}$. 
A good convergence is observed at second order with respect to the combined modes. 
The only third order term present has a rather marginal contribution to the energy with 
respect to $| \Psi_0 \rangle$.
Only after switching to curvilinear coordinates the fluxional motion of the 
highly symmetrical {\zun} cation was correctly accounted for.

The present paper has focused on the definition of an adequate set of coordinates and the
derivation of the expression of the kinetic energy using the polyspherical method. A
convenient way to represent the PES for high-dimensional quantum-dynamical simulation
has been also discussed.
The validity of the Hamiltonian setup has been established by
comparison to available DMC results in the literature. Moreover, the properties of the ground 
vibrational state have been investigated.
Our study provides a picture of the {\zun} system, in which the cluster has to be viewed 
as highly anharmonic, flexible, multi-minima, coupled system. We show that a converged
quantum-dynamical description of such a complex molecular system can still be achieved.
The companion paper following this one focuses on the infrared spectroscopy and dynamics of {\zun}.

\section{Acknowledgments} 

  The authors thank Prof. J. Bowman for providing the potential-energy routine,
  M. Brill for the help with the parallelized code,
  and the Scientific Supercomputing Center Karlsruhe for generously providing computer time.
  O. V. is grateful to the Alexander von Humboldt Foundation for financial support.
  Travel support by the Deutsche Forschungsgemeinschaft (DFG) is also gratefully
  acknowledged.


\clearpage
  \begin{table}
      \caption{Definition of the one-dimensional grids. $N$ denotes the number
      of grid points and $x_i$, $x_f$ the location of first and last point.
      The DVRs are defined in Appendix B of Ref. \cite{bec00:1}.
               }
      \label{tab:1Dgrids}
      \begin{tabular}{c@{\hspace{0.5cm}}c@{\hspace{0.5cm}}c@{\hspace{0.5cm}}c@{\hspace{0.5cm}}l}
    \hline
    \hline
       Coord.              &   $N$  &     $x_i$   &  $x_f$   &  DVR   \\
    \hline
       $z$                 &   27   &     -1.8    &  1.8     &  HO   \\
       $\alpha$            &   21   &      0      &  $2\pi$  &  exp  \\
       $x$                 &    5   &     -0.9    &  0.9     &  HO   \\
       $y$                 &    5   &     -0.9    &  0.9     &  HO   \\
    \hline
       $R$                 &   16   &      4.2    &  6.5     &  HO   \\
       $u_{\beta_{A}}$     &    7   &     -0.5    &  0.5     &  sin \\
       $u_{\beta_{B}}$     &    7   &     -0.5    &  0.5     &  sin \\
    \hline
       $\gamma_A$          &   19   &   $\pi$-1.8 &  $\pi$+1.8 & sin  \\
       $\gamma_B$          &   19   &     -1.8    &  1.8     &  sin  \\
    \hline
       $R_{1 A}$           &    9   &      0.5    &  1.8     &  HO   \\
       $R_{2 A}$           &    9   &      2.2    &  3.8     &  HO   \\
       $u_{\theta_{1 A}}$  &    7   &     -0.5    &  0.5     &  sin \\
    \hline
       $R_{1 B}$           &    9   &      0.5    &  1.8     &  HO   \\
       $R_{2 B}$           &    9   &      2.2    &  3.8     &  HO   \\
       $u_{\theta_{1 B}}$  &    7   &     -0.5    &  0.5     &  sin \\
    \hline
    \hline
      \end{tabular}
  \end{table}

\clearpage  
  \begin{table}[h!]
        \caption{Comparison of the zero point energy (ZPE) of the {\zun} cation
                 calculated by various approaches on the PES
                 by Huang et. al.\cite{hua05:044308}:
                 diffusion Monte-Carlo (DMC), 
                 normal mode analysis (harmonic), 
                 vibrational CI single reference (VCI-SR) and
                 reaction path (VCI-RP) as published in \cite{mc05:061101} and
                 MCTDH results. 
                 $\Delta$ denotes the difference to the DMC result. The
                 converged MCTDH result is obtained with $10\,500\,000$ configurations. 
                 Compare with Table \ref{tab:zpe-mctdh}.
                 }
        \label{tab:zpe}
        \begin{center}
        \begin{tabular}{ccc}
           \hline
           \hline
       \hspace{1.5cm}Method\hspace{1.5cm}  &  ZPE(\icm) & $\Delta$(\icm)  \\
           \hline
           DMC                & $12\,393$    & $0$    \\ 
           harmonic                & $12\,635$    & $242$  \\
           VCI-SR             & $12\,590$    & $197$  \\
           VCI-RP             & $12\,497$    & $104$  \\
           MCTDH             & $12\,376.3$    & $-16.7$ \\
           \hline
           \hline
        \end{tabular}
        \end{center}
       \end{table}

\clearpage       
         \begin{table}[h]
         \caption{Comparison of the zero point energy (ZPE) of the {\zun} cation
                  between different MCTDH calculations with ascending number of
                  configurations. The $\Delta$ values are given with respect to
                  the diffusion Monte Carlo result, $12\,393$ {\icm} \cite{mc05:061101}.
                  }
         \label{tab:zpe-mctdh}
         \begin{center}
         \begin{tabular}{c@{\hspace{1cm}}r@{\hspace{1cm}}c@{\hspace{1cm}}c}
            \hline
            \hline
               SPFs per Mode   &  N configs. & ZPE(\icm) &  $\Delta$(\icm)  \\
            \hline
            $(20,20,12,6,6)$   &     $172\,800$        & $12\,383.7$    & $-9.3$    \\ 
            $(35,25,15,8,8)$   &     $840\,000$        & $12\,378.5$    & $-14.5$    \\ 
            $(40,40,20,8,8)$   &  $2\,048\,000$        & $12\,377.8$    & $-15.2$    \\ 
            $(60,40,20,8,8)$   &  $3\,072\,000$        & $12\,376.7$    & $-16.3$    \\ 
            $(70,50,30,10,10)$ & $10\,500\,000$        & $12\,376.3$    & $-16.7$    \\ 
            \hline
            \hline
         \end{tabular}
         \end{center}
        \end{table}

\clearpage
    \begin{table}[hbt]
      \caption{
         Expectation value of the different terms of the potential expansion (central column)
         and square root of the expectation value of the potential squared (right column).
         All energies in {\icm}. The combined modes read:
         $Q_1 = [z,\alpha,x,y]$,
         $Q_2 = [\gamma_A,\gamma_B]$,
         $Q_3 = [R, u_{\beta_{A}}, u_{\beta_{B}}]$,
         $Q_4 = [R_{1 A}, R_{2 A}, u_{\theta_{1 A}}]$ and
         $Q_5 = [R_{1 B}, R_{2 B}, u_{\theta_{1 B}}]$.
         }
      \label{tab:cluExpect}
     \begin{center}
       \begin{tabular}{lcc}
       \hline
       \hline
     &    $\langle\Psi_0|V|\Psi_0\rangle$ &      $\langle\Psi_0|V^2|\Psi_0\rangle^{1/2}$    \\  
      \hline
             $V^{(1)}(Q_1)$                  &    1293.6       &   1807.7              \\
             $V^{(1)}(Q_2)$                  &     750.6       &    966.9              \\
             $V^{(1)}(Q_3)$                  &     171.5       &    266.9              \\
             $V^{(1)}(Q_4)$                  &    2293.2       &   3062.8              \\
             $V^{(1)}(Q_5)$                  &    2293.1       &   3062.8              \\
             $V^{(2)}(Q_1,Q_2)$              &   -526.9        &   1037.2              \\
             $V^{(2)}(Q_1,Q_3)$              &   -78.8         &    290.2              \\
             $V^{(2)}(Q_1,Q_4)$              &   -27.5         &    231.8              \\
             $V^{(2)}(Q_1,Q_4)$              &   -27.4         &    231.7              \\
             $V^{(2)}(Q_2,Q_3)$              &   -10.5         &     37.6              \\
             $V^{(2)}(Q_2,Q_4)$              &   -24.7         &    117.5              \\
             $V^{(2)}(Q_2,Q_5)$              &   -24.7         &    117.9              \\
             $V^{(2)}(Q_3,Q_4)$              &   -18.8         &    180.9              \\
             $V^{(2)}(Q_3,Q_5)$              &   -18.8         &    180.9              \\
             $V^{(2)}(Q_4,Q_5)$              &     1.2         &      9.9              \\
             $V^{(3)}(z,Q_2,Q_3)$            &     1.0         &     50.4              \\
      \hline
      \hline
      \end{tabular}
     \end{center}
    \end{table}

\clearpage
\section*{Figure Captions}

  \figcaption{fig:Vect}{Jacobi description of the H$_5$O$_2^+$ system.
                        The vector $\vec{R}$ connects the two centers of
                        mass of the water monomers. The vector $\vec{r}$
                        connects the center of mass of the water dimer
                        with the central proton.}

  \figcaption{fig:Coord}{Definition of the angles for the H$_5$O$_2^+$ system.
                         The angles $\alpha_A$ and $\alpha_B$ describe the rotation
                         of the water monomers around the vector $\vec{R}$, or equivalently
                         around the z-axis of the E2- or BF-frame.}

  \figcaption{fig:mode-comb}{Number of grid points needed for the representation
              of the clusters of the PES expansion. $m$ is the number of coordinates
              making a mode. 10 grid points per coordinates and 15 coordinates
              are assumed. A horizontal line is drawn at $10^8$, which tentatively
              signals the maximum practical number of points both regarding their
              calculation and the use of the grids in the dynamical calculations.}

  \figcaption{fig:refGeos}{Geometries of the 10 reference points used in the PES expansion.
            The view is along the O-H-O axis. Hence only the closest of
            the two oxygen
            and the four hydrogens can be seen.
            The difference between the geometries in the left column and each
            geometry at the right column is a rotation of $\pi$ along $\alpha$.
            Equivalently, the pairs of structures (a,b), (c,j), (e,h), (g,f), (i,d)
            are related by a permutation of hydrogen atoms of one of the monomers.
            The following coordinates are identical for all reference points:
            $R=4.70$ au,
            $x,y,z=0$,
            $R_{1(A,B)}=1.07$ au,
            $R_{2(A,B)}=2.98$ au,
            $\theta_{1A(1B)}=0$,
            $u_{\beta_{A(B)}}=0$.
            Only coordinates $\alpha$, $\gamma_A$ and $\gamma_B$
            differ at the 10 reference points.
            }

  \figcaption{fig:wavefunc}{For the ground vibrational state probability density along selected coordinates
          and integration over the rest: 
          probability density along the $z$ proton-transfer coordinate (a),
          along the $\alpha$ internal rotation coordinate (b) and
          on the 2D space spanned by the wagging $\gamma_A$ and $\gamma_B$ coordinates (c).
          The dotted line in (b) corresponds to a 10 times enlarged scale.
          It indicates that the probability density at $\alpha=\pi$ is not vanishing.
          }

\clearpage

\clearpage

\clearpage

\clearpage

\clearpage

\end{document}